\documentclass[nojss]{jss}% FOR VIGNETTE
\usepackage{bm}
\usepackage{amsmath}
\usepackage{amssymb}
\usepackage{algorithm}
\usepackage{threeparttable}
\usepackage{algpseudocode}
\usepackage{subcaption}
\usepackage{multirow}
\usepackage{booktabs}
\usepackage{url}
\usepackage{hyperref}
\definecolor{orange}{rgb}{0.93, 0.53, 0.18}
\usepackage{graphicx}
\usepackage{pifont}% http://ctan.org/pkg/pifont
\newcommand{\cmark}{\ding{51}}%
\newcommand{\xmark}{\ding{55}}%

\author{Marta Crispino$^*$\\ Bank of Italy\And Cristina Mollica$^*$\\ %Department of Statistical Sciences
Sapienza University of Rome\newline \And Lucia Modugno \\Bank of Italy\\\\
$^*$ jointly first authors}

\title{\pkg{MSmix}: An \proglang{R} Package for clustering partial rankings via mixtures of Mallows Models with Spearman distance}

\Plainauthor{Marta Crispino, Cristina Mollica, Lucia Modugno} %% comma-separated
\Plaintitle{MSmix: An {R} Package for clustering partial rankings via mixtures of Mallows Model with Spearman distance} %% without formatting
\Shorttitle{MSmix:  Mallows Model with Spearman distance} %% a short title (if necessary)

\Abstract{
  \pkg{MSmix} is a recently developed \proglang{R} package implementing maximum likelihood estimation of finite mixtures of Mallows models with Spearman distance for full and partial rankings. The package is designed to implement computationally tractable estimation routines of the model parameters, with the ability to handle arbitrary forms of partial rankings and sequences of a large number of items. The frequentist estimation task is accomplished via EM algorithms, integrating data augmentation strategies to recover the unobserved heterogeneity and the missing ranks. The package also provides functionalities for uncertainty quantification of the estimated parameters, via diverse bootstrap methods and asymptotic confidence intervals. %The package includes a variety of utilities for data simulation, manipulation, and sample summaries. 
  Generic methods for S3 class objects are constructed for more effectively managing the output of the main routines.
  The usefulness of the package and its computational performance compared with competing software is illustrated via applications to both simulated and original real ranking datasets.
}

\Plainkeywords{mallows model, partial rankings, mixture models, em algorithm, monte carlo, bootstrap} 

\Address{
{Marta Crispino\\
  Department of Economics Statistics and Research\\
  Bank of Italy, Rome, Italy\\
  \email{marta.crispino@bancaditalia.it}}\\\\
  {Cristina Mollica\\
 Department of Statistical Sciences\\
 Sapienza University of Rome, Italy\\
 \email{cristina.mollica@uniroma1.it}}\\\\  {Lucia  Modugno\\
 Department of Economics, Statistics and Research\\
  Bank of Italy, Rome, Italy\\
  \email{lucia.modugno@bancaditalia.it}
  }
  }

\begin{document}

\section[Intro]{Introduction}
Ranking data play a pivotal role in numerous research and practical domains, serving as a means to compare and order a set of $n$ items according to personal preferences or other relevant criteria. From market analysis to sports competitions, from academic assessments to online recommendation systems, rankings are ubiquitous in modern society to capture human choice behaviors or, more generally, ordinal comparison processes in various contexts.

The Mallows model (MM) is a widely-used probabilistic framework for modeling and analyzing ranking data, and is also recognised as a useful parametric tool for rank aggregation tasks \citep{Marden1995}. It is grounded on the assumption that in the population there exists a modal consensus ranking of the $n$ items which best captures the collective preferences. The probability of observing any given ranking decreases as its distance from the consensus increases. Traditionally, the choice of the distance in defining the MM was driven by computational considerations, particularly the availability of the model normalizing constant (or \textit{partition function}) in a closed form. This favored the use of Kendall, Cayley, and Hamming metrics while the Spearman distance has been poorly investigated due to its perceived intractability, despite  representing a meaningful choice in preference domains and rank-based approaches \citep{crispino23efficient}. In fact,  \citet{crispino23efficient} recovered the effectiveness of the Spearman distance in the MM as an adequate metric joining both computational feasibility and interpretability. By leveraging the properties of the Spearman distance, and by means of a novel approximation of the model partition function, the authors addressed the critical inferential challenges that traditionally limited the use of the Spearman distance. This enabled them to propose an efficient strategy to fit the Mallows model with Spearman distance (MMS) with arbitrary forms of partial rankings. Additionally, they extended the model to finite mixtures, allowing to handle the possible unobserved sample heterogeneity.

In the \proglang{R}  environment, few packages implement the MM (or generalizations thereof) for ranking data analysis. 
\pkg{BayesMallows} \citep{BayesMallows} is the unique package adopting the Bayesian perspective to perform inference for the MM and its finite mixture extension.
The flexibility of \pkg{BayesMallows} stands in the wide range of supported distances (including the Spearman) and ranked data formats (complete and partial rankings as well as pairwise comparisons). Moreover, \pkg{BayesMallows} provides estimation uncertainty through the construction of posterior credible sets for the model parameters. Despite Bayesian inference of ranking data is effectively addressed, the  \proglang{R} packages currently available on the Comprehensive R Archive Network (CRAN) provide users with less flexibility and computational performance when considering the frequentist framework. For example, \pkg{pmr} \citep{pmr_R} performs maximum likelihood estimation (MLE) of several ranking models, including the MM with Kendall, Footrule and Spearman distance.  
However, despite the variety of parametric distributions, \pkg{pmr} does not handle neither partial rankings nor mixtures.
Additionally, the estimation routines require the enumeration of all $n!$ permutations for the global search of the consensus ranking MLE and the na\"ive computation of the partition function, implying that the analysis of ranking datasets with $n\geq 12$ items is unfeasible.
The \pkg{rankdist} package \citep{rankdist} fits mixtures of MMs with various basic and weighted metrics, including the Spearman, on a sample of either full or top-$k$ partial rankings through the Expectation-Maximization (EM) algorithm. The implementation related to the use of the Kendall distance is very efficient, whereas, 
similarly to \pkg{pmr}, the partition function of the MMS is roughly coded as the summation over all the $n!$ permutations, and the MLE of the consensus ranking is achieved with a time-consuming neighbor-checking local search.  
Therefore, the procedures can be computationally demanding, especially in the mixture application and, in any case, do not support the analysis of full rankings with $n\geq12$ items or of top-$k$ rankings with $n\geq8$ items. Other packages related to the MM, but limited to the Kendall distance, are \pkg{RMallow} \citep{rmallow}, which fits the MM and mixtures thereof to both full or partially-observed ranking data, and \pkg{ExtMallows} \citep{extmallows}, which supports the MM as well as the extended MM \citep{EMMjasa}. 

Our review underscores that most of the available packages for frequentist estimation of the MM focus on distances admitting a convenient analytical expression of the model normalizing constant (more often, the Kendall), in the attempt to simplify the estimation task. 
Moreover, regardless of the chosen metric, these packages face common limitations, particularly in handling large datasets and partial rankings, typically restricted to top-$k$ sequences. These computational constraints impose restrictions on the sample size, the number of items and the censoring patterns they can feasibly handle. Finally, the current implementations generally lack methods for quantifying MLE uncertainty, particularly for consensus ranking or when a finite mixture is assumed. 

The novel \proglang{R} package \pkg{MSmix} enhances the current suite of methods for mixture-based analysis of partial ranking, enlarging the applicability of finite mixtures of MMS (MMS-mix) to full and partial rankings.  It achieves several methodological and computational advances to overcome the practical limitations experienced with the existing packages, namely: 1) implementation of a recent normalizing constant approximation and of the closed-form MLE of the consensus ranking, to allow inference for the MMS even with a large number of items; 2) analysis of arbitrary forms of partial rankings in the observed sample via data augmentation strategies; 3) availability of routines for measuring estimation uncertainty of all model parameters, through diverse bootstrapping approaches and Hessian-based standard errors; 4) possible parallel execution of the EM algorithms over multiple starting points, to better and more efficiently explore the critical mixed-type parameter space. 

The paper is organized as follows. In Section \ref{sec:background}, we first review the general formulation and estimation algorithms of the MMS and of its finite mixture extension. We then detail the considered approaches for inferential uncertainty quantification. Section \ref{sec:package_arch} outlines the overall package architecture, the main computational aspects, and shows a comparison with existing packages.  Section \ref{sec:format} represents the core part of the paper illustrating the usage of the routines included in \pkg{MSmix}, with applications to brand new ranking datasets and simulations.
Finally, the Section \ref{sec:concl} discusses  possible directions for future releases of our package.

\section{Methodological background}
\label{sec:background}
\subsection{The Mallows model with Spearman distance and its mixture extension}\label{ssec:inference_complete}

Let $\bm{r}=(r_1,\dots,r_n)$ be a ranking of $n$ items, with the generic entry $r_i$ indicating the rank assigned to item $i$.
We adopt the usual convention 
for which $r_i<r_{i'}$ means that item $i$ is preferred to item $i'$ (the lower the rank, the more preferred the item). 
Both items and ranks are identified with the set $\{1,\dots,n\}$, implying that a generic observation $\bm{r}$ is a permutation of the first $n$ integers belonging to the finite discrete space $\mathcal{P}_n$.

The MMS assumes that the probability of observing the ranking $\bm{r}$ is
 \begin{equation*}
 \label{eq:MM}
\mathbb{P}(\bm{r}\,\vert \bm{\rho},\theta)
=\frac{e^{-\theta\, d(\bm{r},\bm\rho)}}{Z(\theta)}
\qquad\qquad\bm{r}\in\mathcal{P}_n,
\end{equation*}

where $\bm\rho\in\mathcal{P}_n$ is the \textit{consensus ranking}, $\theta\in\mathbb{R}_0^+$ is the \textit{concentration} parameter, $d(\bm{r},\bm\rho)=\sum_{i=1}^n(r_i-\rho_i)^2$ is the Spearman distance, and $Z(\theta)=\sum_{\bm{r} \in \mathcal{P}_{n}} e^{-\theta\, d(\bm{r},\bm e)}$, with $\bm e=(1, 2, ..., n)$, is the normalizing constant or \textit{partition function}. 

Let $\underline{\bm{r}}=\{\bm{r}_1,\dots,\bm{r}_N\}$ be a random sample of $N$ rankings drawn from the MMS and $N_l$ be the frequency of the $l$-th distinct observed ranked sequence $\bm{r}_l$,
%for $l=1,\dots,L$
such that $\sum_{l=1}^L N_l=N$. As shown in \cite{crispino23efficient}, the observed-data log-likelihood can be written as follows
\begin{equation}
\begin{split}
\ell(\bm{\rho},\theta\vert\underline{\bm{r}})
=-N\left(\log{Z(\theta)}+2\theta\left(c_n-\bm{\rho}^T{\bm{\bar r}}\right)\right),
\end{split}
\end{equation}
where $c_n=n(n+1)(2n+1)/6$, the symbol $^T$ denotes the transposition (row vector),  ${\bm{\bar r}}=(\bar{r}_1,\ldots,\bar{r}_n)$ is the sample mean rank vector whose $i$-th entry is $\bar{r}_i=\frac{1}{N}\sum_{l=1}^LN_lr_{li}$, and $\bm{\rho}^T{\bm{\bar r}}=\sum_{i=1}^n\rho_i\bar{r}_i$ is the scalar product. The MLE of the consensus ranking is given by the %Borda rank aggregation method, which is the 
ranking arising from ordering the items according to their sample average rank, 
\begin{equation}
    \hat{\bm{\rho}}=(\hat\rho_1,\ldots,\hat\rho_i,\ldots,\hat\rho_n)\quad\text{with}\quad\hat\rho_i=\text{rank}(\bar{\bm r})_i \,,
\end{equation}
 
also known as Borda ranking.
\citet{MurphyMartin2003} showed that the MLE $\hat\theta$ of the concentration parameter is the value equating the expected Spearman distance under the MMS, $\mathbb{E}_\theta(D)$, to the sample average Spearman distance,  $\bar{d}=\frac{1}{N}\sum_{l=1}^LN_ld(\bm{r}_l,\hat{\bm{\rho}})$.
The problem can be easily solved via a root-finding algorithm, provided that one can evaluate the expected Spearman distance, given by
$$\mathbb{E}_{\theta}[D] = \frac{\sum_{\bm{r} \in \mathcal{P}_{n}} d(\bm{r},\bm e) e^{-\theta\, d(\bm{r},\bm e)}}{Z(\theta)}=
\frac{\sum_{d\in\mathcal{D}_n}dN_d \,e^{-d\theta}}{\sum_{d\in\mathcal{D}_n}N_d \,e^{-d\theta}}$$
with $\mathcal{D}_n=\left\{d=2h\, : \, h\in\mathbb{N}_0\text{ and } 0\leq d\leq 2\binom{n+1}{3}\right\}$ and $N_d=\vert\{\bm{r}\in\mathcal{P}_{n}\,:\,d(\bm{r},\bm{e})=d\}\vert$.

%The root of the above equation \eqref{e:thetaest} can be solved via a root-finding algorithm. 
%although the calculations are demanding for all but small values of $n$, because of the intractability of the model normalizing constant. Notice, however, that solving equation \eqref{e:thetaest} requires the knowledge of the frequency distribution of the Spearman distance, because 
%by considering that
%$$-Z'(\theta)/Z(\theta) = 
% $$\mathbb{E}_{\theta}[D] = \frac{\sum_{d\in\mathcal{D}_n}dN_d \,e^{-d\theta}}{\sum_{d\in\mathcal{D}_n}N_d \,e^{-d\theta}}$$
% with $\mathcal{D}_n=\left\{d=2h\, : \, h\in\mathbb{N}_0\text{ and } 0\leq d\leq 2\binom{n+1}{3}\right\}$ and $N_d=\vert\{\bm{r}^*\in\mathcal{P}_{n}\,:\,d(\bm{r}^*,\bm{e})=d\}\vert$. 
The exact values of the frequencies $N_d$ are available for $n\leq 20$. So, in order to tackle inference on rankings of a larger number of items, \cite{crispino23efficient} introduced a novel approximation of the Spearman distance distribution based on Large Deviation theory principles.
In \pkg{MSmix}, we implement their strategy, so that when the normalizing constant and the expected Spearman distance cannot be computed exactly, inference targets an approximation. In Algorithm \ref{alg:full}, we illustrate the steps described above.

\begin{algorithm}\small
\caption{ML estimation of the MMS parameters from full rankings}\label{alg:full}
\hspace*{\algorithmicindent} \textbf{Input}: $\underline{\bm{r}}=\{\bm{r}_1,\dots,\bm{r}_N\}$ observed sample of full $n-$rankings
\begin{enumerate}
    \item[] \textbf{Preliminary steps}: 
\begin{itemize}
    \item[-] For $l=1,\dots,L$, compute the frequency $N_l$ of each distinct $\bm{r}_l$
    \item[-] Compute either the exact or the
approximate frequency distribution of the Spearman distance $\{d,N_d\}_{d\in\mathcal{D}_n}$
    \end{itemize}
    \item Compute the MLE of the consensus ranking $\bm\rho$:
    \begin{enumerate}
        \item Compute the sample mean rank vector ${\bm{\bar r}}=(\bar{r}_1,\ldots,\bar{r}_n)$ 
        \item Compute $\hat{\bm{\rho}}=\text{rank}({\bm{\bar r}})$
    \end{enumerate}
    \item Compute the MLE of the concentration parameter $\theta$:
    \begin{enumerate}
        \item  Compute the sample average distance,  $\bar d=\frac{1}{N}\sum_{l=1}^LN_ld(\bm{r}_l,\hat{\bm{\rho}})=2(c_n-\hat{\bm{\rho}}^T{\bm{\bar r}})$
        \item  Apply \code{uniroot} to find the solution of the equation $\mathbb{E}_{\theta}(D) = 2(c_n-\hat{\bm{\rho}}^T{\bm{\bar r}})$ in $\theta$
    \end{enumerate}
\end{enumerate}
    \hspace*{\algorithmicindent} \textbf{Output}: $\hat{\bm{\rho}}$ and $\hat{\theta}$ 
\end{algorithm}

In order to account for the unobserved sample heterogeneity typical in real ranking data and, more generally, to increase the model flexibility, a MMS-mix is usually adopted. Under the MMS-mix, the sampling distribution is assumed to be 
\begin{equation*}
\label{e:CLog.lik.mixEPL}
\mathbb{P}(\bm{r}|\underline{\bm{\rho}},{\bm{\theta}},{\bm{\omega}})
=\sum_{g=1}^G\omega_g\mathbb{P}(\bm{r}\,|\bm{\rho}_g,\theta_g)
=\sum_{g=1}^G\omega_g\frac{e^{-2\theta_g\, \left(c_n-\bm{\rho}_g^T\bm{r}\right)}}{Z(\theta_g)},
\end{equation*}
with $\omega_g$ and $(\bm{\rho}_g,\theta_g)$ denoting respectively the weight and the pair of MMS parameters of the $g$-th mixture component.

\cite{MurphyMartin2003} first proposed an EM algorithm to fit such mixture models, but
the more efficient version described by \cite{crispino23efficient} is implemented in the \pkg{MSmix} package.
By denoting the latent group membership of the $l-$th distinct observed ranking with $\bm z_l = (z_{l1}, . . . , z_{lG})$, where
$z_{lg} = 1$ if the observation belongs to component $g$ and $z_{lg} = 0$ otherwise, this EM algorithm is sketched in Algorithm \ref{alg:full_mixture}.

\begin{algorithm}\small
\caption{ML estimation of the MMS-mix parameters from full rankings}\label{alg:full_mixture}
\hspace*{\algorithmicindent} \textbf{Input}: $\underline{\bm{r}}=\{\bm{r}_1,\dots,\bm{r}_N\}$ observed sample of full $n-$rankings; $G$ number of clusters

\begin{enumerate}
\item[] \textbf{Preliminary steps}: 
\begin{itemize}
    \item[-] For $l=1,\dots,L$, compute the frequency $N_l$ of each distinct $\bm{r}_l$
    \item[-] Compute either the exact or the
approximate frequency distribution of the Spearman distance $\{d,N_d\}_{d\in\mathcal{D}_n}$
    \end{itemize}
    
    \item[] \textbf{E-step}: for $l=1,\dots,L$ and $g=1,...,G$, compute 

$$ \hat z_{lg} = \frac{\hat\omega_g\mathbb{P}(\bm{r}_l\,|\hat{\bm{\rho}}_g,\hat\theta_g)}{\sum_{g\prime=1}^G\hat\omega_{g\prime}\mathbb{P}(\bm{r}_l\,|\hat{\bm{\rho}}_{g\prime},\hat\theta_{g\prime})}$$

    \item[]\textbf{M-step}: for $g=1,...,G$ compute
    \begin{enumerate}
        \item $\hat \omega_g =\hat{N}_g/N$ with $\hat{N}_g =\sum_{l=1}^L N_l \hat z_{lg}$  
        \item The MLE of $\bm\rho_g$ as in Step 1 of Algorithm \ref{alg:full},  by replacing $\bar{\bm r}$ with\\ 
        $\bar{\bm r}_g = (\bar r_{g1},\dots, \bar r_{gn})$, where $\bar r_{gi} = \frac{1}{\hat N_g}\sum_{l=1}^L N_l\hat z_{lg}r_{li}$
        %{\color{blue}Marta: in realtà qui secondo me non facciamo così ma usiamo hard clustering --> check}
        \item The MLE of $\theta_g$ as in step 2 of Algorithm \ref{alg:full}, by replacing $\bar{\bm r}$ with $\bar{\bm r}_g$ and $\hat{\bm\rho}$ with $\hat{\bm\rho}_g$
    \end{enumerate}
\end{enumerate}
    \hspace*{\algorithmicindent} \textbf{Output}: $\underline{\hat{\bm{\rho}}}=\{\hat{\bm{\rho}}_1,\dots,\hat{\bm{\rho}}_G\},{\hat{\bm\theta}}=\{\hat{\theta}_1,\dots,\hat{\theta}_G\},{\hat{\bm\omega}}=\{\hat\omega_1,\dots,\hat\omega_G\}$, and $\underline{\hat{\bm z}}=\{\bm \hat z_1,\dots,\bm \hat z_N\}$ 
\end{algorithm}

\subsection{Inference on partial rankings}
\label{ssec:partial}
Algorithms \ref{alg:full} and \ref{alg:full_mixture} are very accurate and fast with full rankings, and their implementation effectively supports a large number of items. However, when the sample includes partial sequences, an additional step to handle the missing information is required. 

In \pkg{MSmix}, we implement two schemes to draw inference on partial data. One was recently proposed in \cite{crispino23efficient}, that extend the EM algorithm originally described by \cite{Beckett} to the finite mixture framework, allowing ML inference of the MMS-mix from partial rankings. The key idea is the data augmentation strategy of each distinct partially observed ranking $\bm{r}_l$ with the corresponding set $\mathcal{C}(\bm r_l)\subset \mathcal{P}_n$ of all compatible full rankings.\footnote{This approach obeys to the common \textit{maximum entropy principle}, according to which the possible latent full sequences are equally likely.} 
 
Let $N_m$ be the latent frequency of a distinct full ranking $\bm{r}^*_m\in\cup_{l=1}^L\mathcal{C}(\bm{r}_l)$,
%with 
for which $\sum_{m=1}^MN_m=\vert\cup_{l=1}^L\mathcal{C}(\bm r_l)\vert$. 
The complete-data log-likelihood of the MMS-mix can be written as 
\begin{equation}\label{eq:loglik_compl}
\ell_c(\underline{\bm{\rho}},{\bm{\theta}},{\bm{\omega}},\underline{\bm{z}},\underline{\bm{r}}^*\vert\underline{\bm r})
=\sum_{m=1}^M\sum_{g=1}^GN_mz_{mg}\left(\log\omega_g-2\theta_g\left(c_n-{\bm{\rho}^T_g}\bm{r}^*_m\right)-\log Z(\theta_g)\right). 
\end{equation}
The EM algorithm to maximize \eqref{eq:loglik_compl} by \cite{crispino23efficient} is outlined in Algorithm \ref{alg:partial_mixture}.

\begin{algorithm}\small
\caption{ML estimation of the MMS-mix parameters from partial rankings}\label{alg:partial_mixture}
\hspace*{\algorithmicindent} \textbf{Input}: $\underline{\bm{r}}=\{\bm{r}_1,\dots,\bm{r}_N\}$ observed sample of partial $n-$rankings; $G$ number of clusters

\begin{enumerate}
\item[] \textbf{Preliminary steps}: for $l=1,\dots,L$,
\begin{itemize}
    \item[-] Compute the frequency $N_l$ of each distinct $\bm{r}_l$
    \item[-] Compute and store the sets $\mathcal{C}(\bm r_l)$ of full rankings compatible with each distinct $\bm{r}_l$
\end{itemize} 
    \item[] \textbf{E-step}:
    \begin{enumerate}
        \item For each distinct $\bm r_l$ with $l=1,\dots,L$ and for each $\bm{r}^*_{m^\prime}\in\mathcal{C}(\bm r_l)$, compute 
$$\hat{p}_{lm^\prime}=\mathbb{P}(\bm{r}^*_{m^\prime}\,|\bm{r}_l,\underline{\hat{\bm{\rho}}},{\hat{\bm\theta}},{\hat{\bm\omega}})
=\frac{\sum_{g=1}^G \hat\omega_g e^{-2\hat\theta_g\left(c_n-{\hat{\bm{\rho}}_g^T}\bm{r}^*_{m^\prime}\right)-\log Z\left(\hat\theta_g\right)}}{\sum_{\bm{s}^*\in \mathcal{C}(\bm{r}_l)}\sum_{g=1}^G \hat\omega_ge^{-2\hat\theta_g\left(c_n-\hat{\bm{\rho}}^T_{g}\bm{s}^*\right)-\log Z\left(\hat\theta_g\right)}}
$$
\item For $m=1,\dots,M$, compute
$
\hat{N}_m=\sum_{l:\,\bm{r}^*_{m^\prime}\in\mathcal{C}(\bm{r}_l)} N_l \hat{p}_{lm^\prime}
$
        \item For $m=1,\dots,M$, and $g=1,\dots,G$, compute
$$\hat z_{mg}=\dfrac{\hat\omega_g\mathbb{P}\big(\bm{r}^*_m\big\vert\hat{\bm{\rho}}_g,\hat\theta_g\big)}{\sum_{g'=1}^G\hat\omega_{g'}\mathbb{P}\big(\bm{r}^*_m\big\vert\hat{\bm{\rho}}_{g'},\hat\theta_{g'}\big)}$$

    \end{enumerate}

    \item[] \textbf{M-step}: for $g=1,\dots,G$, compute
    \begin{itemize}
        \item[-] $\hat\omega_g =\hat{N}_g/N$ with $\hat{N}_g=\sum_{m=1}^M\hat{N}_m\hat z_{mg}$
        \item[-] The MLE of $\bm\rho_g$ as in M-step (b) of Algorithm \ref{alg:full_mixture},  by replacing $\bar{\bm r}_g$ with\\
        $\bar{\bm r}^*_g = (\bar r^*_{g1},\dots, \bar r^*_{gn})$, where $\bar r^*_{gi} = \frac{1}{\hat N_g}\sum_{m=1}^M \hat N_m \hat z_{mg}r^*_{mi}$
        \item[-] The MLE of $\theta_g$ as in M-step (c) of Algorithm \ref{alg:full_mixture}, by substituting $\bar{\bm r}_g$ with $\bar{\bm r}^*_g$ 
    \end{itemize}
\end{enumerate}
   \hspace*{\algorithmicindent} \textbf{Output}: $\underline{\hat{\bm{\rho}}}=\{\hat{\bm{\rho}}_1,\dots,\hat{\bm{\rho}}_G\},{\hat{\bm\theta}}=\{\hat{\theta}_1,\dots,\hat{\theta}_G\},{\hat{\bm\omega}}=\{\hat\omega_1,\dots,\hat\omega_G\}\text{ and }\underline{\hat{\bm z}}=\{\bm \hat z_1,\dots,\bm \hat z_N\}$ 
\end{algorithm}

Even if effective from the inferential point of view, Algorithm \ref{alg:partial_mixture} can be computationally intensive and demands for a lot of memory with many censored positions (larger than 10, say) and large sample sizes.
This happens because the data augmentation step requires the preliminary construction and iterative computations on the list of the possibly large sets $\mathcal{C}(\bm r_l)$ associated to each partial observation. 
To address this issue, in \pkg{MSmix} we propose the use of a Monte Carlo (MC) step in the EM algorithm (MCEM) \citep{wei_tanner} as an additional inferential procedure for the MMS-mix in case of partial rankings.
The core idea is to iteratively complete the missing ranks by sampling from the postulated MMS-mix conditionally on the current parameter values (MC-step).

Let $\mathcal{I}_{s}\subset \{1,2,\dots,n\}$ be the subset of %$n_j=|\mathcal{I}_{j}|$ 
items actually ranked in the observed partial ranking $\bm{r}_s$.\footnote{Note that for a better account of sampling variability and exploration of the parameter space, the MCEM algorithm works at the level of the single observed units, indexed by $s=1,\dots,N$,  instead of the aggregated data $(\bm r_l, N_l)_{l=1,\dots,L}$. } %and $\bm{r}^*_m\in\mathcal{C}(\bm{r}_l)$ one of its $(N-n_l)!$ compatible full rankings such that $r_{li}=r_{mi}$ for $i\in\mathcal{I}_{l}$. 
The MC step is designed as follows:
\begin{description}
\item[MC-step:] for $\bm{r}_s$, $s=1,\dots,N$, simulate
\begin{align}
\tilde{\bm{z}}_s\,\vert\,\hat{\bm{z}}_s&\sim\text{Multinom}\big(1,(\hat z_{s1},\dots,\hat z_{sG})\big) \label{eq:mcem1}\\
\tilde{\bm{r}}_s\,\vert\,\underline{\bm{\rho}},{\bm{\theta}},\tilde{\bm{z}}_s&\sim\sum_{g=1}^G\tilde{z}_{sg}\mathbb{P}\left(\bm{r}|\bm{\rho}_{g},\kappa\theta_{g}\right)
\label{eq:mcem1bis}\end{align}
and complete the partial ranking $\bm{r}_s$ with the full sequence $\bm{r}^*_s=(r^*_{s1},\dots,r^*_{sn})$ such that $r^*_{si}=r_{si}$ for $i\in\mathcal{I}_s$ whereas,
for $i\notin\mathcal{I}_s$, the positions must be assigned to the items so that their relative ranks match those in $\tilde{\bm{r}}_s$.
\end{description}

The value $\kappa > 0$ in equation \eqref{eq:mcem1bis}
 is a tuning constant that affects the precision of the sampled rankings in the MC step. Essentially, the tuning serves to possibly increase the variability ($0<\kappa<1$) or the concentration ($\kappa >1$) of the sampled rankings around the current consensus ranking.  %This option can prove useful to control some blablA, especially when there are many missing positions.
 The MCEM scheme is detailed in Algorithm \ref{alg:partial_mcem}.

\begin{algorithm}\small
\caption{ML estimation of the MMS-mix parameters from partial rankings (MCEM)}\label{alg:partial_mcem}
\hspace*{\algorithmicindent} \textbf{Input}: $\underline{\bm{r}}=\{\bm{r}_1,\dots,\bm{r}_N\}$ observed sample of partial $n-$rankings; $G$ number of clusters

\begin{enumerate}
\item[] \textbf{Preliminary step}: for $s=1,\dots,N$, complete $\bm{r}_s$ at random, obtaining a full ranking $\bm{r}^*_s\in \mathcal{C}(\bm r_s)$
   
    \item[] \textbf{E-step}: for $s=1,\dots,N$  compute $\hat{\bm z}_s = (\hat{z}_{s1}, \dots , \hat{z}_{sG})$ as in E-step of Algorithm \ref{alg:full_mixture}, by replacing $\bm r_l$ with $\bm r^*_s$
     
 \item[] \textbf{M-step}: same as in Algorithm \ref{alg:full_mixture}

 \item[] \textbf{MC-step}: for $s=1,\dots,N$, complete $\bm{r}_s$  with the scheme \eqref{eq:mcem1}-\eqref{eq:mcem1bis}, obtaining an update for $\bm{r}^*_s\in\mathcal{C}(\bm r_s)$ 
\end{enumerate}
   \hspace*{\algorithmicindent} \textbf{Output}: $\underline{\hat{\bm{\rho}}}=\{\hat{\bm{\rho}}_1,\dots,\hat{\bm{\rho}}_G\},{\hat{\bm\theta}}=\{\hat{\theta}_1,\dots,\hat{\theta}_G\},{\hat{\bm\omega}}=\{\hat\omega_1,\dots,\hat\omega_G\}$, and $\underline{\hat{\bm z}}=\{\bm \hat z_1,\dots,\bm \hat z_N\}$ 
\end{algorithm}

Both augmentation schemes have been generalized and optimized to work effectively across a spectrum of censoring patterns, rather than being limited solely to the top$-k$ scenario. 

\subsection{Estimation uncertainty}
\label{subsec:confint}
\subsubsection{Asymptotic confidence intervals}
\label{subsubsec:confint_hess}
To quantify estimation uncertainty, we constructed confidence sets based on the asymptotic likelihood theory and bootstrap procedures.

The former approach was adopted for the continuous parameters (i.e., precision and weights). Specifically, when the consensus ranking $\bm\rho_g$ is assumed to be known, the asymptotic confidence interval (CI) at level $(1-\alpha)$ for $\hat\theta_g$ is
\begin{equation}\label{ic_theta}
    \left[\hat\theta_g-\frac{\texttt{z}_{1-\alpha/2}}{\sqrt{\hat{N}_g\; \mathbb{V}_{\hat\theta_g}[D]}},\hat\theta_g+\frac{\texttt{z}_{1-\alpha/2}}{\sqrt{\hat{N}_g\; \mathbb{V}_{\hat\theta_g}[D]}}\right]
\end{equation}
where $\alpha\in(0,1)$, $\texttt{z}_{1-\alpha/2}$ is the quantile at level $(1-\alpha/2)$ of the standard normal density and $\mathbb{V}_{\hat\theta_g}[D]$ is the variance of the Spearman distance under the MMS \citep[][Section 6.2]{Marden1995}. The above result follows from the fact that, for a regular and canonical exponential family, $\hat\theta_{\text{MLE}} \approx \mathcal{N}(\theta, \hat{I}^{-1}(\hat\theta))$, where $\hat{I}^{-1}(\hat\theta)$ is the observed Fisher information which, for the MMS,
%the observed Fisher information 
is equal to $\mathbb{V}_{\hat\theta_g}[D]$ \citep[see][Supplementary material]{crispino2022informative}. When $\bm\rho_g$ is unknown, the regularity conditions of the MMS likelihood do no longer hold, due to the presence of the discrete component $\mathcal{P}_n$ in the parameter space. However, \cite{critchlow85metric} proved that, asymptotically, the CI \eqref{ic_theta} still provides a good approximate estimation set for $\theta_g$. 

The standard errors of the mixture weights are determined from the inverse of the observed Fisher information matrix, as described in \citet{mclachan2000}.

\subsubsection{Bootstraped confidence intervals}
\label{subsubsec:confint_hess}

Since the validity of the asymptotic CIs pertains a large sample size approximation, we resort also to a non-parametric bootstrap approach  \citep{efron82boot}.

By indicating with $B$ the total number of bootstrap samples, the steps to be repeated for $b=1,\dots,B$ are the following:
    \begin{itemize}
        \item draw with replacement a sample $\underline{\bm r}^{(b)} =\left\{\bm r_1^{(b)},\dots,\bm r_N^{(b)}\right\}$ from the observed data $\underline{\bm r}$.\footnote{
For full rankings and a single mixture component, the \pkg{MSmix} package also offers the parametric bootstrap method, where each simulated sample $\underline{\bm r}^{(b)}$ is obtained by randomly sampling from the fitted MMS rather then from the observed data.}
        \item compute the MLE
        %$\bar{\bm{r}}^{(b)}$ and 
        $\hat{\bm\rho}^{(b)}$
        % , i.e. %the sample mean rank vector and  
        % the MLE of the parameter $\bm\rho$ based 
        on the $b$-th bootstrap sample $\underline{\bm r}^{(b)}$.
        \item compute the MLE %$\bar{\bm{r}}^{(b)}$ and 
        $\hat{\theta}^{(b)}$ %, i.e. %the sample mean rank vector and  
        %the MLE of the parameter $\theta$ based 
        on the $b$-th bootstrap sample $\underline{\bm r}^{(b)}$.
    \end{itemize}
The resulting sequences $\{\hat{\bm\rho}^{(1)},\dots,\hat{\bm\rho}^{(B)}\}$ and $\{\hat{\theta}^{(1)},\dots,\hat{\theta}^{(B)}\}$ of $B$  bootstrap MLEs serve to estimate the sampling variability of $\hat{\bm\rho}$ and $\hat{\theta}$ respectively. 

To summarize the uncertainty on the discrete consensus parameter, we construct itemwise CIs, providing plausible sets of ranks separately for each item. To guarantee narrower intervals as well as a proper account of possible multimodality, these are obtained as highest probability regions of the $n$ bootstrap first-order marginals, that is the sets of most likely ranks for each items at the given $100(1-\alpha)\%$ level of confidence.
We also provide a way to visualize the variability of the bootstrap MLEs through a heatmap of the first-order marginals,
% matrix of the bootstrap MLEs, 
that is the $n\times n$ matrix whose $(i,j)-$th element is given by 
$\frac{1}{B}\sum_{b=1}^{B} \mathbb{I}_{[\hat{\rho}^{(b)}_i=j]}.$

For the continuous concentration parameter, the bounds of the $100(1-\alpha)\%$ CIs can be determined as the quantiles at level $\alpha/2$ and $(1 - \alpha/2)$ of the MLE bootstrap sample.

In the presence of multiple mixture components ($G>1$), the CIs of the component-specific parameters are determined using the non-parametric bootstrap method applied on each subsample of rankings allocated to the $G$ clusters \citep{taushanov2019bootstrap}. We considered two approaches to perform this allocation: i) the deterministic MAP classification (\textit{separated method}) or ii) a simulated classification at each iteration $b$ from a multinomial distribution with the estimated posterior membership probabilities $\underline{\hat{\bm{z}}}$ (\textit{soft method}). The key difference between the two methods is that the separated one ignores the uncertainty in cluster assignment, hence, it does not return CIs for the mixture weights and, in general, leads to narrower CIs for the component-specific parameters. In contrast, the soft method accounts for this uncertainty, allowing to construct also intervals for the mixture weights and providing more conservative CIs.

%Table \ref{tab:bootmeth} summaries all the bootstrap methods implemented in the package by type of ranked observations and number of mixture components.

% \begin{table}[h!]
% \centering
% \caption{Bootstrap methods for determining CIs in \pkg{MSmix}}
% \begin{tabular}{lccccc}
% & \multicolumn{2}{c}{$G=1$}  & & \multicolumn{2}{c}{$G>1$}  \\ 
% \cline{2-3}
% \cline{5-6}
% & parametric & non-parametric& & separated & soft  \\ 
% \hline
% Full rankings & \checkmark & \checkmark & & \checkmark &\checkmark \\
% \hline
% Partial ranking &   & \checkmark & & \checkmark & \checkmark \\ 
% \hline
% \end{tabular}
% \label{tab:bootmeth}
% \end{table}

% \begin{table}[b]
% \centering
% \caption{Bootstrap methods for determining CIs in \pkg{MSmix}.}
% \begin{tabular}{lllll}
%                           & \textbf{Full rankings}                                                       & \textbf{Partial rankings}                                         &  &  \\ \cline{1-3}
% $G=1$              & \begin{tabular}[c]{@{}l@{}}non-parametric\\ parametric\end{tabular} & non-parametric                                           &  &  \\ \cline{1-3}
% $G>1$ & \begin{tabular}[c]{@{}l@{}}soft\\ separated\end{tabular}            & \begin{tabular}[c]{@{}l@{}}soft\\ separated\end{tabular} &  &  \\ \cline{1-3}
%                           &                                                                     &                                                          &  & 
% \end{tabular}
% \label{tab:bootmeth}
% \end{table}

\section{Package architecture and implementation}
\label{sec:package_arch}
The \pkg{MSmix} package is available from the CRAN at
\url{https://cran.r-project.org/web/packages/MSmix}. The software is mainly written in \proglang{R} language, but several strategies have been designed to effectively address the computational challenges, especially related to the analysis of large samples of partial rankings with a wide set of alternatives. The key approaches adopted to limit execution time and memory usage are described below.

\begin{itemize}
    \item Even though the input ranking dataset is required in non-aggregated form, as detailed in Section \ref{subsec:format}, most of the proposed inferential algorithms first determine the frequency distribution of the observations, and then work at aggregated level. This step reduces data volume and, consequently, the overall computational burden.
    \item For very large $n$, the approximate Spearman distance distribution is evaluated over a predefined grid of distance values. This approach prevents the computation of frequencies $N_d$ from becoming numerically intractable or prohibitive, both in terms of computational time and memory allocation.
    \item The ranking spaces $\mathcal{P}_n$ for $n\leq 11$, needed for the data augmentation of partial rankings in Algorithm \ref{alg:partial_mixture}, are internally stored in the package and available for an offline use. 
    \item \pkg{MSmix} is one of the few \proglang{R} packages for ranking data which includes the parallelization option of the iterative estimation procedures over multiple initialization. This is crucial to guarantee a good parameter space exploration and convergence achievement at significantly reduced costs in terms of execution time.  \item The implementation of some critical steps is optimized with a call to functions coded in the \proglang{C++} language, such as the essential computation of the Spearman distance.
\end{itemize}

According to their specific task, the objects contained in \pkg{MSmix} can be grouped into five main categories, 
namely
\begin{description}
\item[Ranking data functions:] objects denoted with the prefix \code{"data\_"} that allow to apply several transformations or summaries to the ranking data.
\item[Model functions:] all the routines aimed at performing a MMS-mix analysis.
\item[Ranking datasets:] objects of class \code{"data.frame"} denoted with the prefix \code{"ranks\_"}, which collect the observed rankings in the first $n$ columns and possible covariates. Most of them are original datasets never analyzed earlier in the literature.
\item[Spearman distance functions:] a series of routines related to the Spearman distance computation and its distributional properties. 
\item[S3 class methods:] generic functions for the S3 class objects associated to the main routines of the package.
\end{description}
In Section  \ref{sec:format}, we extensively describe the  usage of the above objects through applications on simulated and real-world data.

\subsection{Performance benchmarking}
The algorithms developed in \pkg{MSmix} result in impressive gains in overall efficiency compared to existing \proglang{R} packages. We here compare the computational performance of \pkg{MSmix} with the only two competing packages, \pkg{pmr} and \pkg{rankdist}, supporting ML inference on the MMS. Their general characteristics are outlined in Table \ref{tab:compare}, highlighting the greater flexibility of MSmix in handling different forms of partial rankings.

\begin{table}[t]
\caption{Characteristics of the existing \proglang{R} packages for MLE of MMS-mix.}\label{tab:compare}
\centering
\begin{tabular}{rcccccc}
 & \multicolumn{2}{c}{\textbf{\begin{tabular}[c]{@{}c@{}}Full\end{tabular}}} & \multicolumn{2}{c}{\textbf{\begin{tabular}[c]{@{}c@{}}Top partial\end{tabular}}} & \multicolumn{2}{c}{\textbf{\begin{tabular}[c]{@{}c@{}}MAR partial\end{tabular}}} \\ \cline{2-7} 
\multicolumn{1}{l}{}          & $G=1$ & $G>1$ & $G=1$                          & $G>1$                           & $G=1$                           & $G>1$                          \\\hline

\pkg{pmr}& {\color{green}\cmark}& {\color{red}\xmark}& {\color{red}\xmark}& {\color{red}\xmark}& {\color{red}\xmark}  & {\color{red}\xmark}\\ 
\hline
\pkg{rankdist} & {\color{green}\cmark}& {\color{green}\cmark}&{\color{green}\cmark}& {\color{green}\cmark}& {\color{red}\xmark} & {\color{red}\xmark}\\ \hline
\pkg{MSmix}    & {\color{green}\cmark}& {\color{green}\cmark}&{\color{green}\cmark} & {\color{green}\cmark}& {\color{green}\cmark} & {\color{green}\cmark}\\ \hline
\end{tabular}
\end{table}

Table \ref{tab:comp1} reports the execution times for an experiment with full rankings and $G=1$, representing the only case supported by all the three packages. Specifically, we simulated $N=100$ full rankings from the MMS with increasing number of items $n$ and then fitted the true model. 
The comparison shows that \pkg{MSmix} outperforms the other packages in all scenarios and its remarkable speed seems almost not to be impacted by $n$, at least up to $n=20$. This happens because in this case the MLEs are actually available in a one-step procedure, without the need to iterate (nor to locally search). 

\begin{table}[t]
\caption{Comparison between \pkg{MSmix}, \pkg{rankdist} and \pkg{pmr}. Computational times (seconds) of the $G=1$ and full rankings experiment.  Note: \emph{not run} indicates that we did not perform the fit because of the excessive computing time. {\color{red}\xmark} indicates not supported.}\label{tab:comp1}
\centering
\begin{tabular}{cccc}
  \hline
 & \pkg{MSmix} & \pkg{rankdist} & \pkg{pmr} \\ 
  \hline
  $n = 5$ & 0.004 & 0.01 & 0.263 \\ 
  $n = 6$ & 0.004 & 0.028 & 3.955 \\ 
  $n = 7$ & 0.003 & 0.276 & 137.781 \\ 
  $n = 8$ & 0.004 & 2.748 &  \emph{not run}\\ 
  $n = 9$ & 0.004 & 32.1 &  \emph{not run}\\
  $n = 10$ & 0.004 & 538.71 &  \emph{not run}\\ 
  $n = 15$ & 0.004 & {\color{red}\xmark} &  {\color{red}\xmark}\\ 
  $n = 20$ & 0.004 & {\color{red}\xmark} &  {\color{red}\xmark}\\ 
  $n = 50$ & 0.031 & {\color{red}\xmark} &  {\color{red}\xmark}\\ 
  $n=100$ & 0.485  & {\color{red}\xmark} &  {\color{red}\xmark} \\\hline
\end{tabular}
\end{table}

Table \ref{tab:comp2} reports the results of two additional experiments supported only by \pkg{MSmix} and \pkg{rankdist}. The first (left panel) concerns inference of a basic MMS ($G=1$) on top partial rankings: we simulated $N=100$ full rankings of $n=7$ items from the MMS, and then censor them with decreasing number of top-$k$ ranked items. The second (right panel) concerns inference of MMS-mix with full rankings: we simulated $N=100$ full rankings of increasing length $n$  from the MMS-mix with $G=2$ components, and then estimated the true model. 
Again, \pkg{MSmix} turns out to be particularly fast and more efficient when compared to the alternative package. Moreover, the choice of $n=7$ is motivated by the fact that  \pkg{rankdist} only works with a maximum of 7 items in case partial rankings are considered. 

The comparative analysis of this section was performed using R version 4.4.0 on a macOS Monterey 12.7.3 (2.5GHz Intel Core i7 quad-core).

\begin{table}[ht]
\caption{Comparison between \pkg{Msmix} and \pkg{rankdist}. $G=1$ and partial top$-k$ rankings experiment (left); $G>1$ and full rankings experiment (right). Computational times (in seconds) averaged over 100 independent replications. }\label{tab:comp2}
\centering
\begin{tabular}{rrr}
  \hline
 & \pkg{MSmix} & \pkg{rankdist} \\ 
  \hline
$k=5$ & 0.029 & 0.301 \\ 
  $k=4$ & 0.041 & 0.321 \\ 
  $k=3$ & 0.064 & 0.386 \\ 
  $k=2$ & 0.103 & 0.543 \\ 
  $k=1$ & 0.122 & 0.673  \\ 
   \hline
\end{tabular}
\hspace{2cm}
\begin{tabular}{rrr}
  \hline
 & \pkg{MSmix} & \pkg{rankdist} \\ 
 \hline
   $n = 5$ & 0.049 & 0.089 \\ 
  $n = 6$ & 0.035 & 0.185 \\ 
  $n = 7$ & 0.023 & 0.262  \\ 
  $n = 8$ & 0.024 & 0.411  \\ 
  $n = 9$ & 0.018 & 0.612 \\ 
   \hline
\end{tabular}
\vspace{0.5cm}
\end{table}

\section{Using the {MSmix} package}
\label{sec:format}
%use antifragility and beers data

\subsection{Installation and data format}
\label{subsec:format}

The \pkg{MSmix} package can be installed from CRAN and loaded in \proglang{R} with the usual commands
\begin{CodeChunk}
\begin{CodeInput}
R> install.packages("MSmix")
R> library("MSmix")
\end{CodeInput}
\end{CodeChunk}

For a general overview of the \pkg{MSmix} content and a brief recap of the underlying methodology,
%and details on the notation used in the documentation, can be access by the user with
the user can simply run on the console
\begin{CodeChunk}
\begin{CodeInput}
R> help("MSmix-package")
\end{CodeInput}
\end{CodeChunk}
% A vignette is also available to offer a quick tour in the package implementation, and can be accessed with the usual command
% \begin{CodeChunk}
% \begin{CodeInput}
% R> vignette("MSmix-package")
% \end{CodeInput}
% \end{CodeChunk}
%whereas the entire package directory with the source code of the methods can be found on the GitHub repository at \href{https://github.com/jhelvy/logitr}{https://github.com/jhelvy/logitr}.

The knowledge of the data format adopted in a package is, especially for ranked sequences, crucial before safely conducting any ranking data analysis. The \pkg{MSmix} package privileges the ranking data format, which is a natural choice for the MM, and the non-aggregate form, meaning that observations must be provided as an integer $N\times n$ \texttt{matrix} or \texttt{data.frame} with each row representing individual partial rankings. Missing positions must be coded as \code{NA}s and ties are not allowed.

We start the illustration of the main functionalities of \pkg{MSmix} by using a new full ranking dataset contained in the package, called \code{ranks_antifragility}. This dataset, stemming from a 2021 survey on Italian startups during the COVID-19 outbreak, collects rankings of $n=7$ crucial Antifragility features.\footnote{Antifragility properties reflect a company's ability to not only adapt but also improve its activity and grow in response to stressors, volatility and disorders caused by critical and unexpected events.}
Since covariates are also included, the $N=99$ full rankings must be extracted from the first $n=7$ columns as follows
\begin{CodeChunk}
\begin{CodeInput}
R> n <- 7
R> ranks_AF <- ranks_antifragility[, 1:n]
R> str(ranks_AF)
\end{CodeInput}
\begin{CodeOutput}
'data.frame':	99 obs. of  7 variables:
 $ Absorption        : int  4 1 3 4 2 2 1 2 4 4 ...
 $ Redundancy        : int  2 4 4 2 3 1 4 1 3 3 ...
 $ Small_stressors   : int  1 3 1 7 4 6 5 4 6 6 ...
 $ Non_monotonicity  : int  3 2 2 1 1 3 2 5 1 7 ...
 $ Requisite_variety : int  5 7 7 3 7 7 7 3 7 2 ...
 $ Emergence         : int  6 6 6 6 6 5 6 7 2 1 ...
 $ Uncoupling        : int  7 5 5 5 5 4 3 6 5 5 ...
\end{CodeOutput}
\end{CodeChunk}

To facilitate the visualization of the outputs, let us shorten the item labels, and then see the appearance of the rankings provided by the very first three startups

\begin{CodeChunk}
\begin{CodeInput}
R> names(ranks_AF) <- substr(x = names(ranks_AF), start = 1, stop = 3)
R> ranks_AF[1:3, ]
\end{CodeInput}
\begin{CodeOutput}
  Abs Red Sma Non Req Eme Unc
1   4   2   1   3   5   6   7
2   1   4   3   2   7   6   5
3   3   4   1   2   7   6   5
\end{CodeOutput}
\end{CodeChunk}

The switch to the ordering format (and viceversa) can be easily realized with the routine \texttt{data\_conversion}, that has the flexibility to support partial sequences with arbitrary patterns of censoring. Here is the transformation into orderings of the above three full rankings.
\begin{CodeChunk}
\begin{CodeInput}
R> data_conversion(data = ranks_AF[1:3, ])
\end{CodeInput}
\begin{CodeOutput}
  [,1] [,2] [,3] [,4] [,5] [,6] [,7]
1    3    2    4    1    5    6    7
2    1    4    3    2    7    6    5
3    3    4    1    2    7    6    5
\end{CodeOutput}
\end{CodeChunk}

\subsection{Data description and manipulation}
\label{subsec:descr}

Descriptive statistics and other useful sample summaries can be obtained with the routine \code{data_description} that, differently from analogous functions supplied by other \proglang{R} packages, can handle partial observations with arbitrary type of missingness. The output is a list of S3 class \code{"data_descr"}, whose components can be displayed with the \code{print.data_descr} method.
For the entire Antifragility sample, the basic application of the command would be
\begin{CodeChunk}
\begin{CodeInput}
R> data_descr_AF <- data_description(rankings = ranks_AF)
R> print(data_descr_AF)
\end{CodeInput}
\begin{CodeOutput}
Sample size: 99 
N. of items: 7 

Frequency distribution of the number of ranked items:

 1  2  3  4  5  6  7 
 0  0  0  0  0  0 99 

Number of missing positions for each item:

Abs Red Sma Non Req Eme Unc 
  0   0   0   0   0   0   0 

Mean rank of each item:

 Abs  Red  Sma  Non  Req  Eme  Unc 
2.45 3.27 4.02 2.71 5.38 5.01 5.15 

Borda ordering:

[1] "Abs" "Non" "Red" "Sma" "Eme" "Unc" "Req"

First-order marginals:

      Abs Red Sma Non Req Eme Unc Sum
Rank1  37  13   6  34   3   3   3  99
Rank2  28  25  10  18   3   9   6  99
Rank3  13  20  22  18  10   7   9  99
Rank4   6  18  28  16  11   9  11  99
Rank5   6  12  14   4  19  20  24  99
Rank6   6   8   9   3  16  39  18  99
Rank7   3   3  10   6  37  12  28  99
Sum    99  99  99  99  99  99  99 693

Pairwise comparison matrix:

    Abs Red Sma Non Req Eme Unc
Abs   0  67  80  52  86  83  82
Red  32   0  63  41  79  79  75
Sma  19  36   0  33  75  68  64
Non  47  58  66   0  86  84  84
Req  13  20  24  13   0  43  47
Eme  16  20  31  15  56   0  59
Unc  17  24  35  15  52  40   0
\end{CodeOutput}
\end{CodeChunk}
where the two displayed matrices correspond, respectively, to:
% i) Frequency distribution of the number of items ranked in each partial sequence.
% \item Mean rank vector.
% \item Borda ordering, given by the vector of the item labels into non-descreasing order of the mean rank vector.
%\item Integer $n\times n$ matrix of 
i) the first-order marginals, with the $(j,i)$-th entry indicating the number of times that item $i$ is ranked in position $j$;
%\item Integer $n\times n$ 
the pairwise comparison matrix, with the $(i,i')$-th entry indicating the number of times that item $i$ is preferred to item $i'$.
%\end{itemize}
The function \code{data_description} also includes an optional \code{subset} argument which allows to summarize specific subsamples defined, for example, through a condition on some of the available covariates. The idea is to facilitate a preliminary exploration of possible different preference patterns influenced by some of the observed subjects' characteristics.

\begin{figure}[t]
      \centering
             \includegraphics[width=\textwidth]{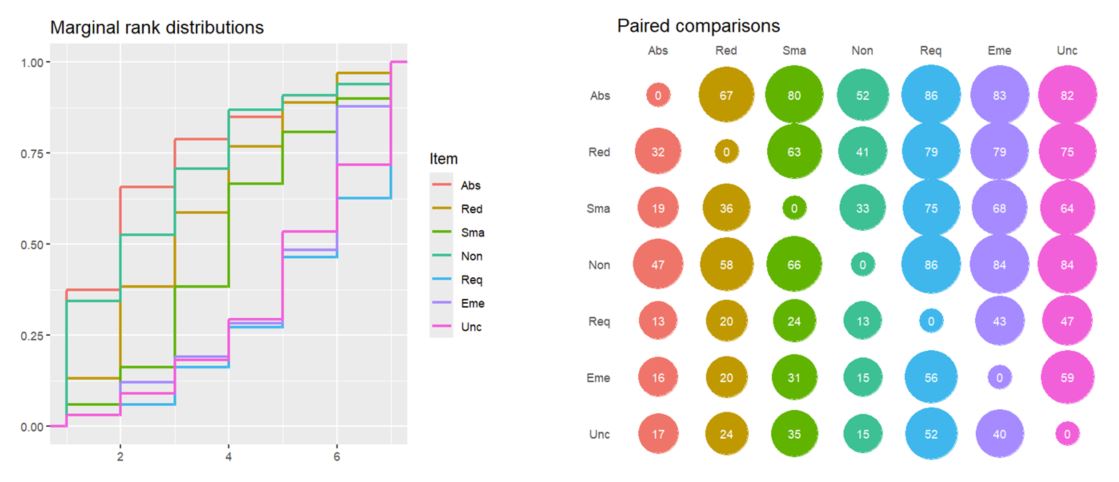}
          \caption{Ecdf's of the marginal rank distributions (left) and bubble plot of the pairwise comparison matrix (right) for the Antifragility dataset.}
        \label{fig:plot.data_descr}
\end{figure}

Finally, we created the \code{plot.data_descr} method to offer a more attractive and intuitive rendering of the fundamental summaries, that is, the simple command
\begin{CodeChunk}
\begin{CodeInput}
R> plot(data_descr_AF)
\end{CodeInput}
\end{CodeChunk}
produces five plots by relying on the fancy graphical tools implemented in the \pkg{ggplot2} package \citep{ggplot}, namely: 1) barplot with the percentages of the number of ranked items, 2) pictogram of the mean rank vector, 3) heatmap of the first-order marginals (either by item or by rank), 4) ecdf's of the marginal rank distributions and 5) bubble plot of the pairwise comparison matrix. For the Antifragility dataset, the plots 4) and 5) are shown in Figure \ref{fig:plot.data_descr}.

% \begin{figure}[t]
%       \centering
%       \begin{subfigure}[t]{0.85\textwidth}
%          \centering
%          \includegraphics[width=\textwidth]{ecdf.pdf}
%          \end{subfigure}
%          \vspace{0.5cm}
       
%        \begin{subfigure}[t]{0.4\textwidth}
%          \centering
%          \includegraphics[width=\textwidth]{bubble.pdf}
%          \end{subfigure}
%         \caption{Heatmap of the first-order marginals (top-left); Ecdf's of the marginal rank distributions (top-right) and bubble plot of the pairwise comparison matrix (bottom) for the Antifragility dataset.}
%         \label{fig:plot.data_descr}
% \end{figure}

Concerning ranking data manipulation, \pkg{MSmix} provides functions designed to switch from complete to partial sequences, with the routine \code{data_censoring}, or from partial to complete sequences, with the routines \code{data_augmentation} and \code{data_completion}. These functions are particularly useful in simulation scenarios for evaluating the robustness of inferential procedures in recovering the actual data-generating mechanisms under various types and extents of censoring and different data augmentation strategies for handling partial data. 

With \code{data_censoring}, the truncation of complete rankings can be either applied with the top-$k$ scheme, or  the missing at random (MAR) procedure. To retain the information on the top positions, the user needs to set the argument \code{type = "topk"}; instead, to retain information on a random set of positions, she needs to set the argument \code{type = "mar"}. Regardless of the censoring process type, the user can choose the number of positions to be retained for each ranked sequence in: (i) a deterministic way, by specifying an integer vector in the \code{nranked} argument, indicating the desired length of each partial sequence; (ii) a stochastic way, by setting \code{nranked = NULL} (default) and providing in the argument \code{probs} the probabilities for the random number of positions to be kept after censoring, that is, for 1, 2, up to $(n-1)$ ranks.\footnote{Recall that a partial sequence with $(n-1)$ observed entries corresponds to a full ranking.} An example of a deterministic top-$k$ censoring scheme is implemented below to covert the complete Antifragility ranking data into top-3 rankings.
\begin{CodeChunk}
\begin{CodeInput}
R> N <- nrow(ranks_AF)
R> top3_AF <- data_censoring(rankings = ranks_AF, type = "topk", 
+                            nranked = rep(3,N))
R> top3_AF$part_rankings[1:3,]
\end{CodeInput}
\end{CodeChunk}
\begin{CodeChunk}
\begin{CodeOutput}
  Abs Red Sma Non Req Eme Unc
1  NA   2   1   3  NA  NA  NA
2   1  NA   3   2  NA  NA  NA
3   3  NA   1   2  NA  NA  NA
\end{CodeOutput}
\end{CodeChunk}
The output of \code{data_censoring} is a list with a first component, named \code{part_rankings}, corresponding to the input complete data matrix \code{rankings} with suitably censored (\code{NA}) entries, and a second component, named \code{nranked}, corresponding to the vector with the number of actually visible positions in each partial ranking.

An example of stochastic top-$k$ censoring scheme on the same dataset, that will result in a random number of bottom positions obscured, can be run as follows 
\begin{CodeChunk}
\begin{CodeInput}
R> top_AF <- data_censoring(rankings = ranks_AF, type = "topk", 
+                           probs = c(1:(n-2),0))
R> top_AF$part_rankings[1:3,]
\end{CodeInput}
\end{CodeChunk}
\begin{CodeChunk}
\begin{CodeOutput}
  Abs Red Sma Non Req Eme Unc
1   4   2   1   3   5  NA  NA
2   1  NA   3   2  NA  NA  NA
3   3  NA   1   2  NA  NA  NA
\end{CodeOutput}
\end{CodeChunk}
\begin{CodeChunk}
\begin{CodeInput}
R> table(top_AF$nranked)
\end{CodeInput}
\end{CodeChunk}
\begin{CodeChunk}
\begin{CodeOutput}
 1  2  3  4  5 
 1 14 16 19 49 
\end{CodeOutput}
\end{CodeChunk}
In this case, the probability vector \code{probs} is assigning an increasing chance of retaining a higher number of top positions, with the exception of a zero value in the last entry, forcing the non-occurrence of full rankings after censoring. Apart from the different setting for the \code{type} argument, applying a MAR censoring scheme requires a similar syntax to the top-$k$. The main difference is that, instead of the censoring process acting only on the bottom part of the rankings, the positions to be censored are determined uniformly at random once the number of ranks to be kept is specified by the user (either deterministically or stochastically).

We conclude this section with an illustration of the counterpart commands of \code{data_censoring} in \pkg{MSmix}, which act on partial rankings and fill in the missing positions with different criteria.
The first, called \code{data_augmentation}, is the key function for estimating MMS-mix on partial rankings via Algorithm \ref{alg:partial_mixture}. Here is a toy example with only two partial rankings characterized by different types of censoring, respectively a top-3 and a MAR sequence
\begin{CodeChunk}
\begin{CodeInput}
R> ranks_toy <- rbind(c(2, NA, 1, NA, 3), c(NA, 4, NA, 1, NA))
R> ranks_toy
\end{CodeInput}
\begin{CodeOutput}
     [,1] [,2] [,3] [,4] [,5]
[1,]    2   NA    1   NA    3
[2,]   NA    4   NA    1   NA
\end{CodeOutput}
\begin{CodeInput}
R> data_augmentation(rankings = ranks_toy)
\end{CodeInput}
\begin{CodeOutput}
[[1]]
     [,1] [,2] [,3] [,4] [,5]
[1,]    2    4    1    5    3
[2,]    2    5    1    4    3

[[2]]
     [,1] [,2] [,3] [,4] [,5]
[1,]    2    4    3    1    5
[2,]    3    4    2    1    5
[3,]    3    4    5    1    2
[4,]    2    4    5    1    3
[5,]    5    4    2    1    3
[6,]    5    4    3    1    2
\end{CodeOutput}
\end{CodeChunk}
The output list contains the matrices of all possible full rankings compatible with each partial sequence.\footnote{These correspond to the sets $\mathcal{C}(\bm r)$ introduced in Section \ref{ssec:partial}.} We remark that, despite the name \code{rankings} of the input partially ranked data matrix, this function can also be applied to partial observations expressed in ordering format. In general, it supports the data augmentation of sequences containing at most 10 missing entries.

The second function, named \code{data_completion}, completes partial rankings with a single full ranking. To complete the rankings in \code{ranks_toy}, one needs to specify the \code{ref_rho} argument with a matrix of the same dimensions as \code{ranks_toy}, containing the reference full rankings in each row. In the example below, we use the identity permutation and its opposite as the reference sequences for completion.
\begin{CodeChunk}
\begin{CodeInput}
R> data_completion(rankings = ranks_toy, ref_rho = rbind(1:5, 5:1)) 
\end{CodeInput}
\end{CodeChunk}
\begin{CodeChunk}
\begin{CodeOutput}
     [,1] [,2] [,3] [,4] [,5]
[1,]    2    4    1    5    3
[2,]    5    4    3    1    2
\end{CodeOutput}
\end{CodeChunk}
The output is the matrix obtained by filling in the missing entries of each partial sequence with the relative positions of the unranked items according to the reference full ranking.\footnote{These sequences correspond to the result of data completion described in the MCEM step of Section \ref{ssec:partial}.} 
%Canonical choices for the reference complete sequences include the sample Borda ranking, the ranking of the top-1 frequencies or other relevant measures of item dominance. 
%Note that in Algorithm \ref{alg:partial_mcem} this function is iteratively applied by using a i.i.d. sample from the MMS-mix as reference rankings. 
\code{data_completion} accommodates any type of censoring, similar to \code{data_augmentation}, but without the need to enumerate all possible orders of missing positions. Consequently, there is no upper limit on the number of \code{NA} entries in the partial sequences. 

\subsection{Sampling}
\label{subsec:sampling}
The function devoted to simulating an i.i.d. sample of full rankings from a MMS-mix is \code{rMSmix}, which relies on the Metropolis-Hastings (MH) procedure implemented in the \proglang{R} package \pkg{BayesMallows} \citep{BayesMallows}. When $n\leq 10$, we also offer the possibility to perform exact sampling. This can be achieved by setting the logical \code{mh} argument to \code{FALSE}.

The \code{rMSmix} function requires the user to specify: i) the desired number of permutations  (\code{sample_size}), ii) the number of items (\code{n_items}) and iii) the number of mixture components (\code{n_clust}). The mixture parameters can be passed with the separated (and optional) arguments \code{rho}, \code{theta} and \code{weights}, set to \code{NULL} by default. If the user does not input the above parameters, the concentrations are sampled uniformly in the interval $(1/n^2,3/n^{3/2})$,\footnote{The concentration parameters play a delicate role. In fact, if $\theta$ is too close to zero, the MMS turns out to be indistinguishable from the uniform distribution on $\mathcal{P}_n$, while if $\theta$  is too large the MMS probability distribution would tend to a Dirac on the consensus ranking $\bm\rho$. The critical magnitude turns out to be $\theta\sim c/n^2$ with $c > 0$ fixed \citep{zhong2021mallows}.} while the simulation of the consensus parameters and the weights can be selected with the logical argument \code{uniform}. The option \code{uniform = TRUE} consists in generating the non-specified parameters uniformly in their support. Here is an example where $N=100$ full rankings of $n=8$ items are exactly generated from a $3-$component MMS-mix, with assigned and equal concentrations $\bm\theta=(.15,.15,.15)$ and the other parameters sampled uniformly at random. 
\begin{CodeChunk}
\begin{CodeInput}
R> theta = rep(.15, 3)
R> sam <- rMSmix(sample_size = 100, n_items = 8, n_clust = 3, theta = theta,
+               uniform = TRUE, mh = FALSE)
\end{CodeInput}
\end{CodeChunk} 

The function \code{rMSmix} returns a list of five named objects: the $N\times n$ matrix with the simulated complete rankings (\code{samples}), the model parameters actually used for the simulation (\code{rho}, \code{theta} and \code{weights}) and the simulated group membership labels (\code{classification}).

For the previous example, they can be extracted as follows
\begin{CodeChunk}
\begin{CodeOutput}
R> sam$samples[1:3,]

     [,1] [,2] [,3] [,4] [,5] [,6] [,7] [,8]
[1,]    6    1    7    5    8    3    2    4
[2,]    2    1    3    7    5    4    8    6
[3,]    6    2    7    5    8    3    1    4

R> sam$rho

     [,1] [,2] [,3] [,4] [,5] [,6] [,7] [,8]
[1,]    6    2    1    5    4    3    8    7
[2,]    4    2    5    3    8    7    1    6
[3,]    6    2    8    4    7    3    1    5

R> sam$weights

[1] 0.49165535 0.04123627 0.46710838

R> table(sam$classification)

 1  2  3 
35  5 60
\end{CodeOutput}
\end{CodeChunk}
One can note that, with uniform sampling, cluster separation and balance of the drawings among the mixture components are not guaranteed. In fact, cluster 2 has a very small weight ($\omega_2 \approx 0.04$) corresponding to  only 5 observations; moreover, the consensus rankings of clusters 2 and 3 are quite similar, as testified by their low relative Spearman distance.\footnote{The maximum Spearman distance among two rankings of length $n$ is given by $2\binom{n+1}{3}$.} 
\begin{CodeChunk}
\begin{CodeInput}
R> max_spear_dist <- 2*choose(8+1,3)
R> spear_dist(rankings = sam$rho[2,], rho = sam$rho[3,])/max_spear_dist
\end{CodeInput}
\begin{CodeOutput}
[1] 0.1904762
\end{CodeOutput}
\end{CodeChunk}

To ensure separation among the mixture components and non-sparse weights, the user can set the option \code{uniform = FALSE}. Specifically, the consensus rankings are drawn with a minimum Spearman distance from each other equal to $\frac{2}{G}\binom{n+1}{3}$, and the mixing weights are sampled from a symmetric Dirichlet distribution with (large) shape parameters $\bm\alpha=(2G,\dots,2G)$ to favour populated and balanced clusters. 
\begin{CodeChunk}
\begin{CodeInput}
R> sam <- rMSmix(sample_size = 100, n_items = 8, n_clust = 3, theta = theta,
+               uniform = FALSE, mh = FALSE)
\end{CodeInput}
\end{CodeChunk}

We notice that now the three clusters are more balanced, and with central rankings at larger relative distance.

\begin{CodeChunk}
\begin{CodeInput}
R> sam$weights
\end{CodeInput}
\begin{CodeOutput}
    [1] 0.5214495 0.2594782 0.2190723
\end{CodeOutput}
\begin{CodeInput}
R> spear_dist(rankings = sam$rho)/max_spear_dist
\end{CodeInput}
\begin{CodeOutput}
              1         2
2 0.6309524          
3 0.7023810 0.6666667
\end{CodeOutput}
\end{CodeChunk}

\begin{figure}[t]
     \centering
     \begin{subfigure}[b]{0.49\textwidth}
         \centering
         \includegraphics[width=\textwidth]{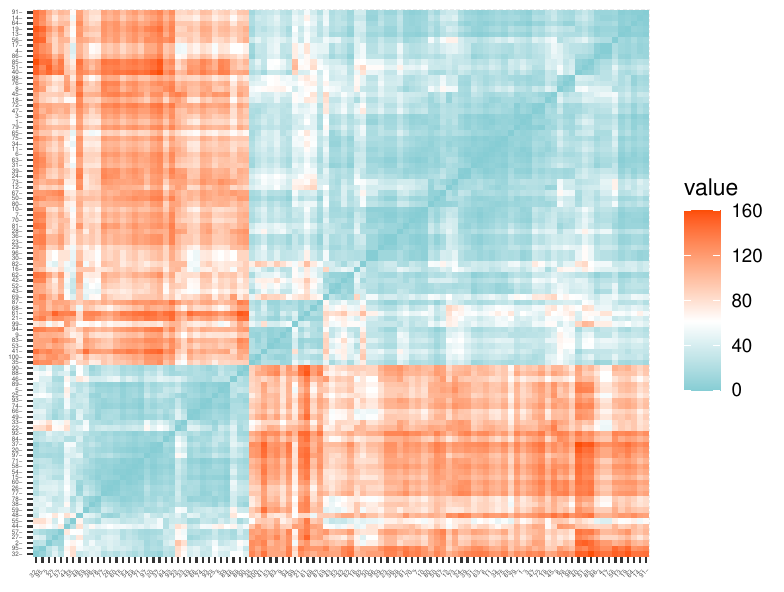}
         \caption{Uniformly sampled parameters.}
         \label{fig:s1a}
     \end{subfigure}
     \hfill
     \begin{subfigure}[b]{0.49\textwidth}
         \centering
         \includegraphics[width=\textwidth]{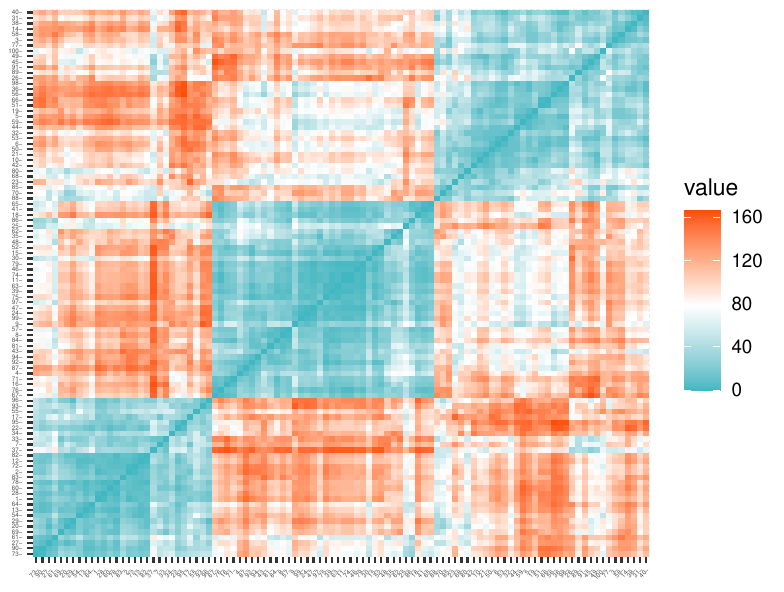}
         \caption{Separated clusters.}
         \label{fig:s1b}
     \end{subfigure}
        \caption{A visualization of the random draws from a MMS-mix, obtained by setting \code{uniform = TRUE} (a), and \code{uniform = FALSE} (b).}
        \label{fig:s1}
\end{figure}

In Figure \ref{fig:s1} we show the separation among clusters in the two examples through the Spearman distance matrix of the simulated samples, which quantifies the dissimilarity between each pair of observations. For example, Figure \ref{fig:s1b} can be constructed as follows\footnote{Notably, the \code{plot.dist} function of \pkg{MSmix} fills in the gap of a generic method for objects of class \code{"dist"} in \proglang{R}, since it allows to visualize, and hence compare, distance matrices of any metric.}
\begin{CodeChunk}
\begin{CodeInput}
R> plot.dist(spear_dist(rankings = sam$samples))
\end{CodeInput}
\end{CodeChunk}

It suggests the presence of only two well-separated clusters in the sample, while three clusters are evident from  Figure \ref{fig:s1b}. 

In conclusion, \code{rMSmix} is designed to facilitate the implementation of alternative sampling schemes, that can be fruitful to assess the performance of the inferential procedures and their robustness under a variety of simulation scenarios.

\subsection{Application on full rankings}
\label{subsec:est_full}
In this section, we show how to perform a mixture model analysis on the Antifragility rankings.
To this aim, we use the command \code{fitMSmix}, the core function of the \pkg{MSmix} package, which performs MLE of the MMS-mix on the input \code{rankings} via EM algorithm with the desired number \code{n_clust} of components. The number of multiple starting points, needed to address the issue of local maxima, can be set through the argument \code{n_start}, and the list \code{init} possibly allows to configure initial values of the parameters for each starting point. 

We now estimate several MMS-mix with a number of components ranging from 1 to 6 and save the BIC (Bayesian informative criterion) values in a separate vector for then choosing the optimal number of clusters.
\begin{CodeChunk}
\begin{CodeInput}
R> FIT.try <- list()
R> BIC <- setNames(numeric(6), paste0('G = ', 1:6))
R> for(i in 1:6){
+    FIT.try[[i]] <- fitMSmix(rankings = ranks_AF, n_clust = i, n_start = 50)
+    BIC[i] <- FIT.try[[i]]$mod$bic}
\end{CodeInput}
\end{CodeChunk}
The BIC values of the six estimated models are
\begin{CodeChunk}
\begin{CodeInput}
R> print(BIC)
   G = 1    G = 2    G = 3    G = 4    G = 5    G = 6 
1494.435 1461.494 1442.749 1444.223 1449.714 1453.101  
\end{CodeInput}
\end{CodeChunk}

\begin{figure}[t]
\centering
\includegraphics[width=0.8\textwidth]{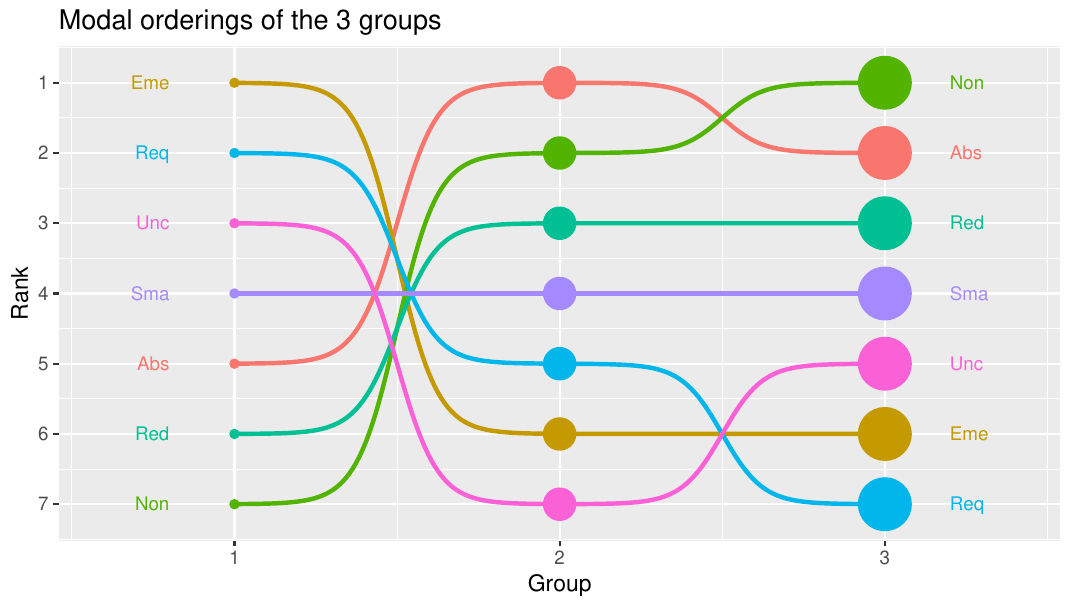}
\caption{Bump plot depicting the estimated consensus rankings of the three clusters for the \code{ranks\_antifragility} dataset, obtained with the generic function \code{plot.emMSmix}.}
 \label{f:antifr_plot1}
\end{figure}
suggesting \code{G = 3} as the optimal number of groups (lowest BIC). 
The function \code{fitMSmix} creates an object of S3 class \code{"emMSmix"}, which is a list whose main component, named  \code{mod}, describes the best fitted model over the \code{n_start} initializations. It includes, for example, the MLE of the parameters (\code{rho}, \code{theta} and \code{weights}), fitting measures (\code{log_lik} and \code{bic}), the estimated posterior membership probabilities (\code{z_hat}) and the related MAP allocation (\code{map_classification}) as well as the indicator of convergence achievement (\code{conv}).  

The MLEs of the best fitted model can be shown also through the \code{summary.emMSmix} method,
\begin{CodeChunk}
\begin{CodeInput}
R> summary(object = FIT.try[[3]])
\end{CodeInput}
\begin{CodeOutput}
Call:
fitMSmix(rankings = ranks_AF, n_clust = 3, n_start = 50)

-----------------------------
--- MLE of the parameters ---
-----------------------------

Component-specific consensus rankings:
       Abs Red Sma Non Req Eme Unc
Group1   5   6   4   7   2   1   3
Group2   1   3   4   2   5   6   7
Group3   2   3   4   1   7   6   5

Component-specific consensus orderings:
       Rank1 Rank2 Rank3 Rank4 Rank5 Rank6 Rank7
Group1 "Eme" "Req" "Unc" "Sma" "Abs" "Red" "Non"
Group2 "Abs" "Non" "Red" "Sma" "Req" "Eme" "Unc"
Group3 "Non" "Abs" "Red" "Sma" "Unc" "Eme" "Req"

Component-specific precisions:
Group1 Group2 Group3 
 0.111  0.241  0.087 

Mixture weights:
Group1 Group2 Group3 
 0.083  0.343  0.574 
\end{CodeOutput}
\end{CodeChunk}
which also displays the estimated modal orderings. The generic function \code{plot.emMSmix} is also associated to the class \code{"emMSmix"} and constructs two fancy plots. The first one is the bump plot (Figure \ref{f:antifr_plot1}) depicting the consensus ranking of each cluster, with different colors assigned to each item, circle sizes proportional to the estimated weights and lines to better highlight item positions in the modal orderings of the various components. For this example, we note that the size of the second cluster is almost half that of the third cluster, while the first cluster is very small. Moreover, the two larger groups (2 and 3) exhibit very similar modal rankings and quite opposite preferences with respect to the first cluster (items such as ``Emergence'', ``Requisite variety'', and ``Uncoupling'' are ranked at the top in cluster 1, but placed at the bottom in groups 2 and 3). 

\begin{figure}[t]
\centering
\includegraphics[width=\textwidth]{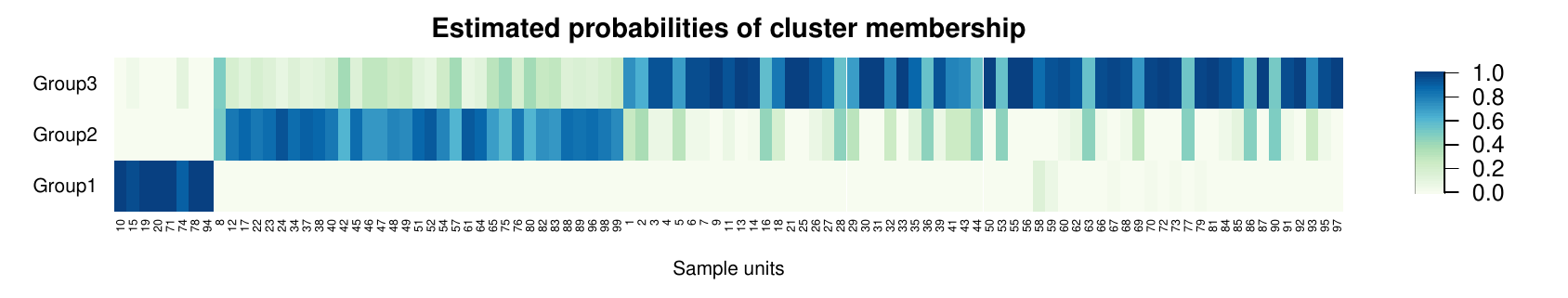}
\caption{Heatplot of the estimated probabilities of cluster memberships for
each observation of the \code{ranks\_antifragility} dataset, obtained with the generic function \code{plot.emMSmix}.}
 \label{f:antifr_plot2}
\end{figure}

Figure \ref{f:antifr_plot2} shows, instead, the individual cluster memberships probabilities, describing the uncertainty with which each observation could be assigned to the mixture components. For example, the units 10, 15, 19, 20, 71, 74, 78 and 94 have high probabilities (close to 1) of belonging to group 1. Instead, some units (e.g., unit 8, 28, 36, and 44) have similar membership probabilities of belonging to clusters 2 or 3, indicating less confidence in their assignment to one of the two groups. On the other hand, when some clusters are close on the ranking space, a certain degree of uncertainty in recovering the true membership is expected.

The package provides also routines for CI computation, working with the object of class \code{"emMSmix"} as first input argument. For example, we can produce Hessian-based CIs for the precisions and mixture weights with \code{confintMSmix}, which is a function specific for full ranking data. With the default confedence level (\code{conf_level = 0.95}), one obtains

\begin{CodeChunk}
\begin{CodeInput}
R> confintMSmix(object = FIT.try[[3]])
\end{CodeInput}
\begin{CodeOutput}
Hessian-based 95%CIs for the precisions:

       lower upper
Group1 0.055 0.168
Group2 0.186 0.296
Group3 0.069 0.105


Hessian-based 95%CIs for the mixture weights:

       lower upper
Group1 0.064 0.101
Group2 0.328 0.359
Group3 0.560 0.588
\end{CodeOutput}
\end{CodeChunk}

Another possibility relies on bootstrap CI calculation. Let us opt for the soft bootstrap method (the default choice when $G>1$) which, unlike the separated one (\code{type = "separated"}), produces CIs also for weights. We require \code{n_boot = 500} bootstrap samples and then print the output object of class \code{"bootMSmix"} through the generic function \code{print.bootMSmix}.

\begin{CodeChunk}
    \begin{CodeInput}
R> CI_bootSoft <- bootstrapMSmix(object = FIT.try[[3]], n_boot = 500, all = TRUE)
R> print(CI_bootSoft)
    \end{CodeInput}
    \begin{CodeOutput}
Bootstrap itemwise 95%CIs for the consensus rankings:

       Abs           Red         Sma           Non     Req         Eme    
Group1 "{3,4,5,6,7}" "{4,5,6,7}" "{2,3,4,5,6}" "{6,7}" "{1,2,3,4}" "{1,2}"
Group2 "{1}"         "{2,3}"     "{4}"         "{2,3}" "{5}"       "{6}"  
Group3 "{1,2,3}"     "{2,3}"     "{4,5}"       "{1,2}" "{7}"       "{5,6}"
       Unc        
Group1 "{2,3,4,5}"
Group2 "{7}"      
Group3 "{4,5,6}"  


Bootstrap 95%CIs for the precisions:

       lower upper
Group1 0.068 0.212
Group2 0.193 0.314
Group3 0.069 0.112


Bootstrap 95%CIs for the mixture weights:

       lower upper
Group1 0.071 0.101
Group2 0.283 0.404
Group3 0.505 0.636
\end{CodeOutput}
\end{CodeChunk}
The bootstrap itemwise intervals for the consensus ranking are wider in the first group, the smallest one, while the second cluster shows very little uncertainty. Note also that the Hessian-based intervals for the precisions and the weights are narrower than the bootstrap ones, except for the weight of the first cluster that has a few observations.

\begin{figure}[t]
     \centering
     \begin{subfigure}[b]{0.49\textwidth}
         \centering
         \includegraphics[width=\textwidth]{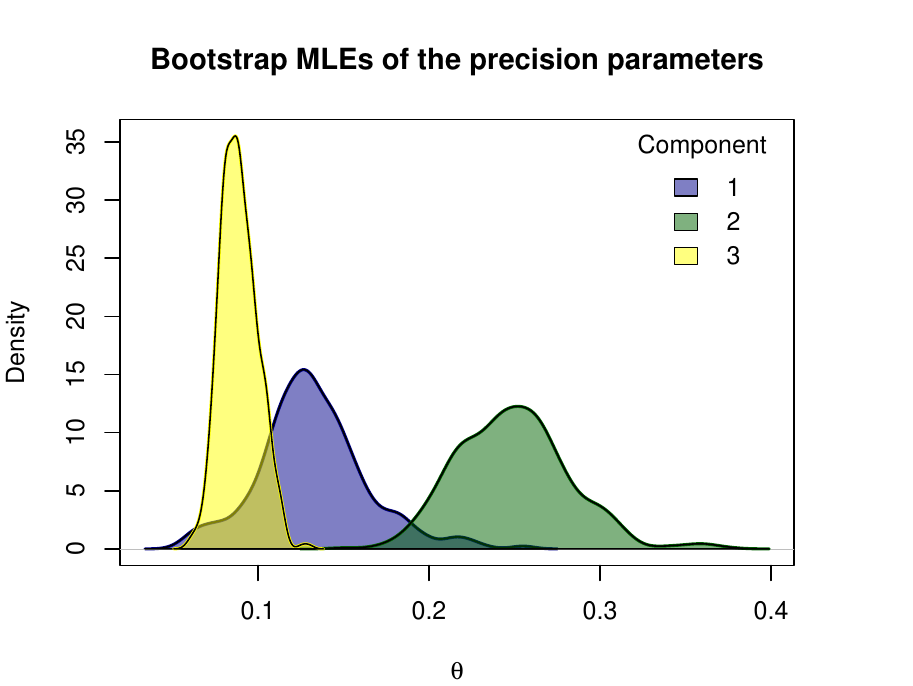}
     \end{subfigure}
 %    \hfill
     \begin{subfigure}[b]{0.49\textwidth}
         \centering
         \includegraphics[width=\textwidth]{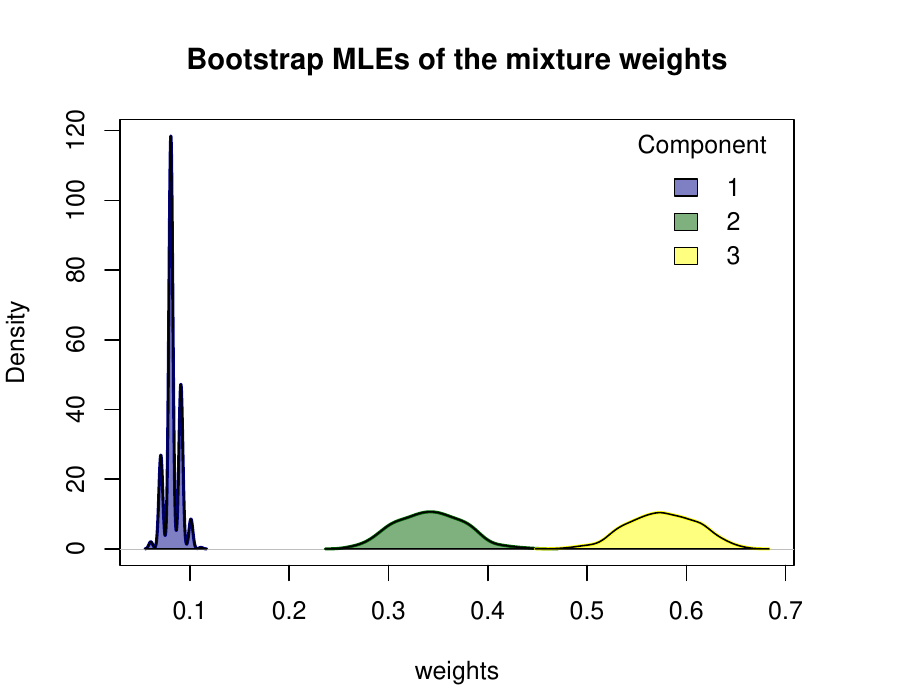}
      \end{subfigure}
       
        \caption{Kernel densities of the soft bootstrap MLEs of the precision parameters (left) and weights (right) for the \code{ranks\_antifragility} dataset.}
        \label{fig:boot_theta_weights}
\end{figure}

The logical argument \code{all} indicates whether the MLEs estimates obtained from the  bootstrap samples must be returned in the output. When \code{all = TRUE}, as in this case, the user can visualize the bootstrap sample variability with the generic function \code{plot.bootMSmix}. It returns the heatmap of the first-order marginals of the bootstrap samples (an example is available in Figure~\ref{fig:heat_beers}), and kernel densities for the precisions and weights (Figure~\ref{fig:boot_theta_weights}).

\subsection{Application on partial rankings}
\label{subsec:est_partial}

In this section we illustrate how to perform inference on the MAR partial rankings by exploiting the original \code{ranks_beers} dataset. 
These data were collected through an online survey administered to the participants of the 2018 Pint of Science festival held in Grenoble. A sample of $N = 105$ subjects provided their partial rankings of $n=20$ beers according to their personal tastes. The rankings are recorded in the first 20 columns of the dataset, while column 21 contains a covariate regarding respondents' residency. 

\begin{figure}[t]
     \centering
     \includegraphics[width=0.65\textwidth]{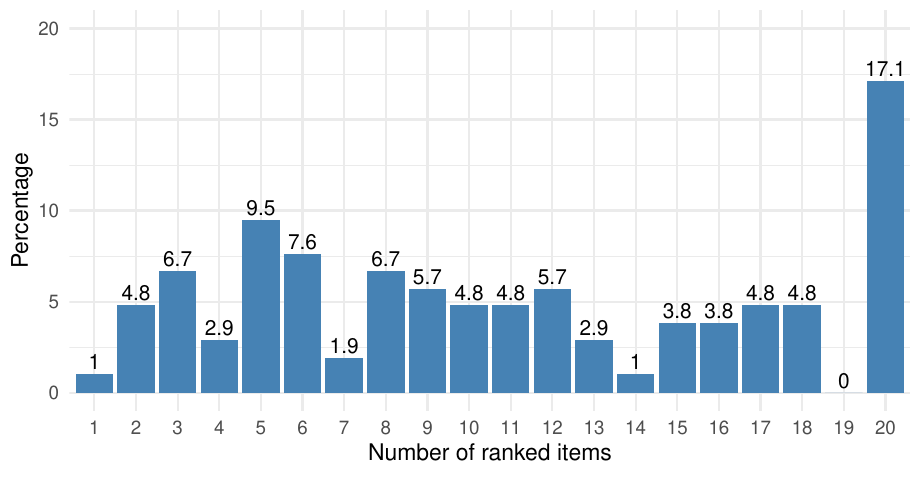}
      \caption{Percentages of the number of ranked items in the \code{ranks\_beers} dataset. This barplot was obtained as first plot in the output of the \code{data\_description} routine.}
     \label{fig:dewcr_beers}
\end{figure}

The barplot with the percentages of the number of beers actually ranked by the participants is reported in Figure \ref{fig:dewcr_beers}.  We restrict the analysis to partial rankings with maximum 8 missing positions, to show both the data augmentation schemes (Algorithms \ref{alg:partial_mixture} and \ref{alg:partial_mcem}) implemented in the package.
Thanks to the \code{subset} argument of \code{fitMSmix}, we can specify the subsample of observations to be considered directly in the fit command.
To speed up the estimation process, we parallelize the multiple starting points by setting  \code{parallel = TRUE}.\footnote{Note that exact reproducibility of this section may not be possible due to the use of parallelization, which can lead to minor variations in inferential results between runs.}

\begin{CodeChunk}
\begin{CodeInput}
R> rankings <- ranks_beers[,1:20]
R> subset_beers <- (rowSums(is.na(rankings)) <= 8)
R> library(doParallel)
R> registerDoParallel(cores = detectCores())
R> FIT_aug <- fitMSmix(rankings,n_clust = 1, n_start = 15, 
+                     subset = subset_beers, mc_em = FALSE, parallel = TRUE)
R> FIT_mcem <- fitMSmix(rankings, n_clust = 1, n_start = 15, 
+                      subset = subset_beers, mc_em = TRUE, parallel = TRUE)
\end{CodeInput}
\end{CodeChunk}

The logical \code{mc_em} argument indicates whether the MCEM scheme (Algorithm \ref{alg:partial_mcem}) must be applied. When \code{mc_em = FALSE} (default),  Algorithm \ref{alg:partial_mixture} is implemented.\footnote{
This type of data augmentation is supported for up to 10 missing positions in the partial rankings. However, it is important to note that while this operation may be feasible in principle for some datasets, it can be slow and memory-intensive. For instance, augmenting and storing all rankings compatible with the subset of the beers dataset with a maximum of 10 missing positions requires more than 3GB of storage space.} 

We note that, for this application, the results of the two methods are very similar.

\begin{CodeChunk}
\begin{CodeInput}
R> spear_dist(FIT_aug$mod$rho,FIT_mcem$mod$rho)/(2+choose(20+1,3)) 

[1] 0.006006006

R> c('theta_aug' = FIT_aug$mod$theta, 'theta_mcem' = FIT_mcem$mod$theta)
  theta_aug  theta_mcem 
0.008580397 0.008964391
\end{CodeInput}
\end{CodeChunk}

One can then evaluate the uncertainty associated to the consensus ranking estimated via the MCEM with the non-parametric bootstrap (default for $G=1$). Also in this case, we can parallelize over the multiple starting points of the EM algorithm used to fit the bootstrap samples.
\begin{CodeChunk}
\begin{CodeInput}
R> boot_mcem <- bootstrapMSmix(object = FIT_mcem, n_boot = 300, n_start = 15, 
                             all = TRUE, parallel = TRUE)

R> print(boot_mcem)

Bootstrap itemwise 95%CIs for the consensus rankings:

       Stella                   Kwak          KronKron  Faro                     
Group1 "{12,13,14,15,16,17,18}" "{2,3,4,5,6}" "{19,20}" "{8,9,10,11,12,13,14,15}"
       Kron1664           Chimay  Pelforth           KronCarls               
Group1 "{14,15,16,17,18}" "{1,2}" "{11,12,13,14,15}" "{12,13,14,15,16,17,18}"
       KronKanter Hoegaarden           Grimbergen      Pietra                 
Group1 "{19,20}"  "{6,7,8,9,10,11,12}" "{2,3,4,5,6,7}" "{6,7,8,9,10,11,12,13}"
       Affligem         Goudale           Leffe             Heineken          
Group1 "{3,4,5,6,7,8}" "{4,5,6,7,8,9,10}" "{6,7,8,9,10,11}" "{14,15,16,17,18}"
       Duvel             Choulette                Orval                       
Group1 "{2,3,4,5,6,7,8}" "{12,13,14,15,16,17,18}" "{5,6,7,8,9,10,11,12,13,15}"
       Karmeliet      
Group1 "{1,2,3,4,5,6}"


Bootstrap 95%CIs for the precisions:

       lower upper
Group1 0.007 0.013

\end{CodeInput}
\end{CodeChunk}

\begin{figure}[t]
     \centering
     \includegraphics[scale=0.8,width=0.9\textwidth]{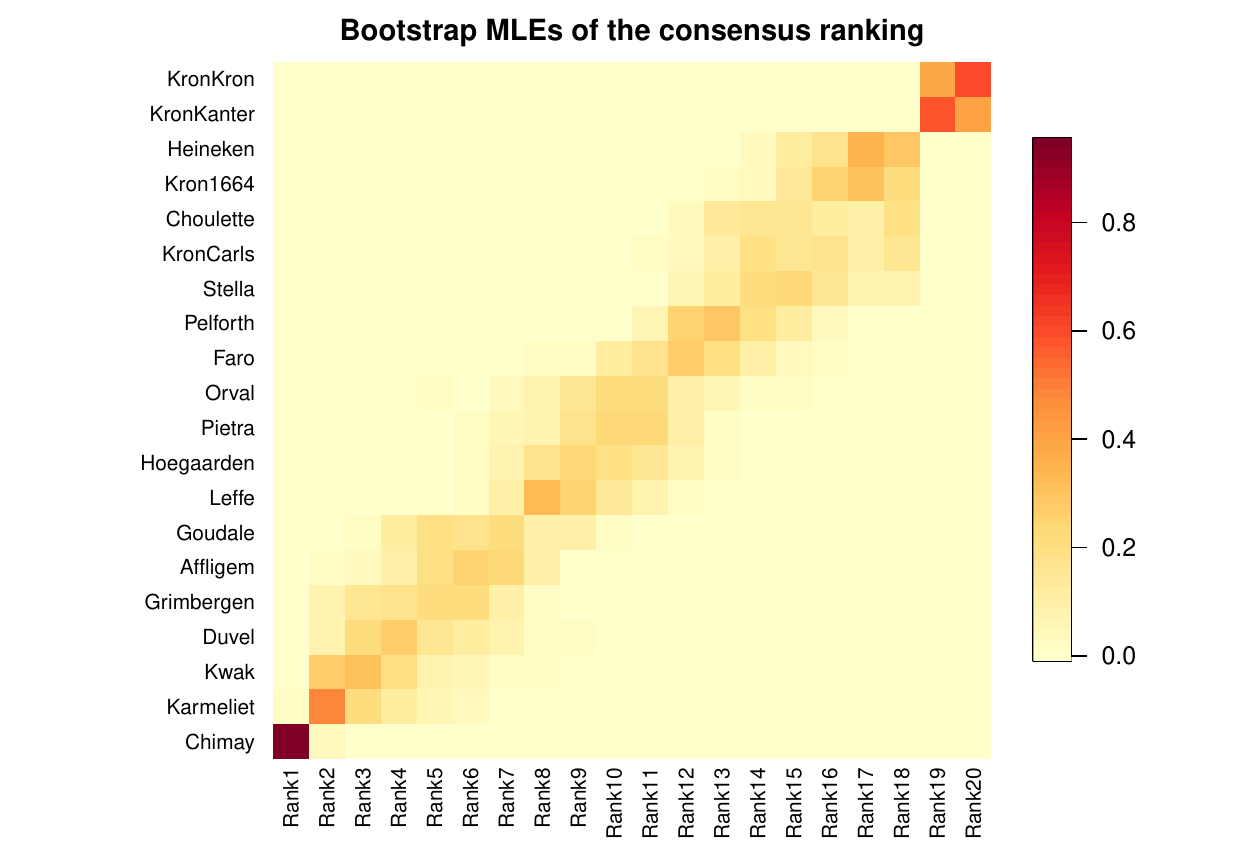}
      \caption{Heatmap of the bootstrap MLE of the consensus ranking for a subsample of the \code{ranks\_beers} dataset. In the y-axis items are ordered according to the MLE of $\bm\rho$ (top-ranked beer at the bottom).}
     \label{fig:heat_beers}
\end{figure}

The plot of the bootstrap output (\code{plot(boot_mcem)}), displayed in Figure \ref{fig:heat_beers}, helps in understanding the variability and confidence in the rankings of the beers. In fact, the top ranked beer (Chimay) and the two bottom ranked ones (KronKanter and KronKron) are quite reliably ranked in those positions. On the contrary, the exact ranks of the other beers are more uncertain, with itemwise  95\% bootstrap-based CIs for some beers being as wide as 10 positions (out of 20). Note also that some itemwise regions can result in subsets of non-contiguous ranks, as in the case of \code{Orval} whose CI does not include rank 14.

We conclude this section by stressing that the application to the beers dataset represents a non-trivial case of ranking data analysis, since currently there are no other \proglang{R} packages supporting MLE on MAR partially-ranked sequences.

\subsection{Additional options}

In the \pkg{MSmix} package there are also some functions to deal with the distribution of the Spearman distance, which are
needed to fit the model (see Section \ref{sec:background}) but that may be useful for external use in their own right. 

The function \code{spear_dist_distr} returns the (log-)frequency distribution of the Spearman distance under the uniform model. If $n\leq 20$, the function returns the exact distribution by relying on a call to the \code{get_cardinalities} routine of \pkg{BayesMallows}. Here is an example
\begin{CodeChunk}
\begin{CodeInput}
R> n_items = 5
R> spear_dist_distr(n_items)

$distances
 [1]  0  2  4  6  8 10 12 14 16 18 20 22 24 26 28 30 32 34 36 38 40

$logcard
 [1] 0.000000 1.386294 1.098612 1.791759 1.945910 1.791759 1.386294 2.302585
 [9] 1.791759 2.302585 1.791759 2.302585 1.791759 2.302585 1.386294 1.791759
[17] 1.945910 1.791759 1.098612 1.386294 0.000000
\end{CodeInput}
\end{CodeChunk}
When $n> 20$, the approximate distribution introduced by \cite{crispino23efficient} is returned and, in the case $n\geq 170$, its calculation is restricted over a fixed grid of values of the Spearman distance to limit computational burden.

The functions \code{partition_fun_spear},  \code{expected_spear_dist} and \code{var_spear_dist} provide, respectively, the partition function $Z(\theta)$, the expected value $\mathbb{E}_{\theta}[D]$ and the variance $\mathbb{V}_{\theta}[D]$ of the Spearman distance under the MMS. For $n=5$, one has
\begin{CodeChunk}
\begin{CodeInput}
R> partition_fun_spear(theta = 0.1, n_items = n_items)
[1] 3.253889
R> expected_spear_dist(theta = 0.1, n_items = n_items)
[1] 2.421115
R> var_spear_dist(theta = 0.1, n_items = n_items)
[1] 4.202741
\end{CodeInput}
\end{CodeChunk}
For these functions, the computation is exact or approximate according to the same principle described for \code{spear_dist_distr}.

\section{Conclusions}\label{sec:concl}
The new \pkg{MSmix} package enriches the \proglang{R} software environment with functions to analyze finite mixtures of MMS on full and partial rankings with arbitrary patterns of censoring. Inference is conducted within the ML framework via EM algorithms. Estimation uncertainty is quantified with bootstrap methods and approximate CIs from the asymptotic likelihood theory. 

The innovative contributions of \pkg{MSmix} span from both methodological and computational advancements to address the lacks and limitations found in most of the existing packages, especially the possibility of realizing a ranking data analysis with many items and missing positions or assessing estimation uncertainty of model parameters. The package also exploits the construction of \texttt{S3} class objects and related generic methods to offer a unified and original analysis framework. In this regard, a special attention was devoted to the development of effective visualization tools and summaries, that can assist the users in the reporting results and designing conclusions with a more transparent account of the associated uncertainty. 

The package architecture is designed to facilitate code extensibility for accomplishing future research directions. For instance, its flexibility in accommodating diverse data censoring patterns could be beneficial for integrating the parametric mixture model with specifications of missing data mechanisms. Moreover, the package capability to analyze data characterized by a large number of alternatives could motivate the interest in clustering similar items, as recently proposed for the MM in \citet{piancastelli}, or even in developing methods to solve bi-clustering problems.  Finally, to better characterize choice processes, the EM algorithms could be integrated with an additional step for estimating the impact of individual and/or item-specific covariates - a typical but complex task in preference analysis from ranking data \citep[see e.g.,][]{ gormley2008mixture,zhu21partition}. 
We are currently working in this direction with a proposal to enrich the MMS-mix with a \textit{Mixture of Experts} model \citep{jacobs,jordan}, that is, a mixture model in which the parameters are functions of the covariates \citep{crispinoMoE}.  Future releases of \pkg{MSmix} will also include functions to deal with different distances among rankings.

\section{Acknowledgments}
The authors wish to thank Prof. Luca Tardella and Prof. Enrico Casadio Tarabusi for the insightful discussions on various aspects of the methodology used in this paper. Additionally, the author wish to thank Prof Maria Vincenza Ciasullo, Dr. Nicola Cucari, Dr. Raffaella Montera, Prof Maria Iannario and Dr. Rosaria Simone for generously sharing their data. 

\bibliography{references}

\end{document}